\documentclass[11pt]{JHEP3}
\DeclareMathAlphabet{\scr}{U}{rsfs}{m}{n}
\usepackage{latexsym}
\usepackage[mathscr]{eucal}
\usepackage{amsfonts}
\usepackage{amscd}
\usepackage{cite}
\usepackage{amssymb}
\usepackage[centertags]{amsmath}
\usepackage{dsfont}
\usepackage{yfonts}
\usepackage{enumerate}
\usepackage{epsfig}
\usepackage{amsmath}
\usepackage{graphicx}
\usepackage{bbold}
\usepackage{bbm}
\usepackage[english]{babel}

\newcommand{\cleqn}{\setcounter{equation}{0}}

\def\a{\alpha}

\def\D{\Delta}
\def\e{\varepsilon}

\def\La{\Lambda}

\def\n{\nu}

\def\f{\varphi}

\def\nn{\nonumber}
\def\mtil{\widetilde{m}}

\def\diag{{\rm diag}}

\newcommand{\newc}{\newcommand}
\newc{\be}{\begin{equation}}
\newc{\ee}{\end{equation}}
\newc{\ben}{\begin{equation*}}
\newc{\een}{\end{equation*}}
\newc{\bea}{\begin{eqnarray}}
\newc{\eea}{\end{eqnarray}}
\newc{\bean}{\begin{eqnarray*}}
\newc{\eean}{\end{eqnarray*}}
\newc{\ol}{\overline}
\newc{\wt}{\widetilde}
\newc{\bs}{\boldsymbol}
\newc{\m}{\mathcal}
\newc{\lra}{~\longrightarrow~}

\newc{\VEV}[1]{\langle #1 \rangle}

\newcommand{\dmsol}{\mbox{$\Delta m^2_{\odot}$}}
\newcommand{\dma}{\mbox{$\Delta m^2_{\rm A}$}}
\newcommand{\dmB}{\mbox{$\Delta m^2_{21}$}}
\newcommand{\dmC}{\mbox{$\Delta m^2_{31}$}}
\newcommand{\dmD}{\mbox{$\Delta m^2_{32}$}}
\newcommand{\gsim}{\lower.7ex\hbox{$\;\stackrel{\textstyle>}{\sim}\;$}}
\newcommand{\lsim}{\lower.7ex\hbox{$\;\stackrel{\textstyle<}{\sim}\;$}}

\newcommand{\eV}{\ensuremath{\:\mathrm{eV}}}

\def\PMNS{{\rm PMNS}}

\title{Majorana Phases and Leptogenesis in See-Saw Models with $A_4$ Symmetry}
\author{C.~Hagedorn\\SISSA and INFN-Sezione di Trieste, Trieste I-34014, Italy\\E-mail:\email{hagedorn@sissa.it}}
\author{E.~Molinaro\\SISSA and INFN-Sezione di Trieste, Trieste I-34014, Italy\\E-mail:\email{molinaro@sissa.it}}
\author{S.~T.~Petcov
\footnote{Also at: INRNE, Bulgarian Academy  of Sciences, 
1784 Sofia, Bulgaria.}
\\SISSA and INFN-Sezione di Trieste, Trieste I-34014, 
Italy;\\IPMU, University of Tokyo, Tokyo, Japan\\E-mail:\email{petcov@sissa.it}
}
\abstract{The related issues of Majorana CP violation 
in the lepton sector and leptogenesis are 
investigated in detail in two 
rather generic 
supersymmetric models with type I see-saw mechanism 
of neutrino mass generation and  $A_4$ flavour symmetry,
which naturally lead at leading order 
to tri-bimaximal  neutrino mixing.
The neutrino sector 
in this class of models
is described at leading order 
by just two real parameters and one phase. 
This leads, in particular, to significant 
low energy constraints on the Majorana phases 
$\alpha_{21}$ and $\alpha_{31}$ 
in the PMNS matrix, which play the role 
of leptogenesis CP violating parameters 
in the generation of the baryon asymmetry of the 
Universe. We find that it is possible to generate 
the correct size and sign of the baryon asymmetry 
in both $A_4$ models. The sign of the baryon asymmetry 
is directly related to the signs of $\sin\alpha_{21}$ and/or    
$\sin\alpha_{31}$.}

\preprint{SISSA 48/2009/EP}

\begin{document}

\cleqn
\section{Introduction}

The presence of two large and one small mixing angles in the lepton sector 
\cite{nudata1,nudata2},
\begin{equation}
\sin ^2 \theta_{12} = 0.304 ^{+0.066} _{-0.054}\;,\;\; 
\sin ^2 \theta_{23} = 0.50 ^{+0.17} _{-0.14}\;,\;\; 
\sin^2 \theta_{13} < 0.056\;\;\; (3 \sigma)\;,
\label{mixingangles}
\end{equation}
%
suggests a pattern of neutrino mixing which is remarkably 
similar to the so called ``tri-bimaximal'' (TB) one \cite{hps}. 
Indeed, in the case of TB mixing,  the solar and atmospheric 
neutrino mixing angles $\theta_{12}$ and $\theta_{23}$ 
have values very close to, or coinciding with, the best fit ones in 
eq. (\ref{mixingangles}), determined in global analyses 
of neutrino oscillation data, 
\begin{equation}
\sin ^2 \theta_{12} = 1/3 \; , \;\;
\sin ^2 \theta_{23} = 1/2 
\;\;\;. 
\end{equation}
The TB mixing scheme predicts also that $\theta_{13}=0$. Correspondingly, 
the 
neutrino mixing matrix is given by
\begin{equation}
U_{\nu} = U_{TB}\,\diag\left(1,e^{i\a_{21}/2},
e^{i\a_{31}/2}\right) \label{U_matrix}
\end{equation}
where 
\begin{equation}
\label{eq:UTB}
U_{TB} = \left(
\begin{array}{ccc}
	\sqrt{2/3}  & 1/\sqrt{3} & 0\\
	-1/\sqrt{6} & 1/\sqrt{3} & 1/\sqrt{2}\\
	-1/\sqrt{6} & 1/\sqrt{3} & -1/\sqrt{2}
\end{array}
\right)
\end{equation}
%
and $\a_{21}$ and $\a_{31}$ are Majorana 
CP violating phases \cite{BHP80,SchVal80}.

  TB mixing can appear naturally in models using the
tetrahedral group $A_4$ as flavor symmetry \cite{A4first}. 
\footnote{It can also be derived from models with an $S_4$ flavor
symmetry \cite{S4}.}
The three generations of left-handed leptons and right-handed (RH)
neutrinos are unified into a triplet representation of
the $A_4$ group, whereas the right-handed charged leptons are
$A_4$-singlets. To be concrete, in the following we focus 
on two models which 
represent a class of $A_4$ models,
namely the models of Altarelli-Feruglio (AF) \cite{AF2} and 
Altarelli-Meloni (AM) \cite{AM}. Both models are based 
on the Standard Model
(SM) gauge symmetry group and are supersymmetric. 
Additional degrees of freedom, flavons, are
introduced in order to appropriately break the $A_4$ flavor symmetry at high
energies. Both models have in common that they
predict at leading order (LO) a diagonal mass matrix 
for charged leptons and lead to exact TB mixing in the neutrino
sector. The mass matrix of the RH neutrinos contains only two
complex parameters $X$, $Z$. Light neutrino masses are 
generated through the type I see-saw mechanism. 
All low energy observables are expressed through
only three independent quantities: $\alpha=|3 Z/X|$, the relative
phase $\phi$ between $X$ and $Z$, and the absolute scale of the light 
neutrino masses. The latter is a combination of the unique 
neutrino Yukawa coupling  and $|X|$ which determines 
the scale of RH neutrino masses. 
The main difference at LO between the AF and AM models 
is in the generation of the 
charged lepton mass hierarchy: in \cite{AF2} an 
additional Froggatt-Nielsen (FN) symmetry $U(1)_{FN}$
is invoked, whereas in \cite{AM} the hierarchy between 
the masses of the charged leptons arises through
multi-flavon insertions 
\footnote{This is due to a different choice in 
the AM model of the vacuum alignment of the flavon, $\varphi_T$,
slightly different transformation properties of the 
right-handed charged leptons under $A_4$
and the presence of a different additional cyclic symmetry.}. 
As a result the next-to-leading (NLO)
corrections in these models are different.
The expansion parameter in the $A_4$ models of interest 
is the vacuum expectation value
(VEV) of a generic flavon field divided by 
the cut-off scale $\La$ of the
theory. Its typical size is $\lambda_c^2 \approx 0.04$ with 
$\lambda_c \approx 0.22$ being of the size of 
the Cabibbo angle.

  The fact that the properties of light as well as 
heavy neutrinos are essentially 
fixed by the three parameters $\alpha$, $\phi$ and $|X|$ leads to strong
constraints. We will be interested, in particular, 
in the dependence of the Majorana CP violating phases in the
Pontecorvo-Maki-Nakagawa-Sakata (PMNS) matrix \cite{BPont57},
which are relevant in leptogenesis and neutrinoless double beta decay,
on $\alpha$ and $\phi$. Actually, $\alpha$ and $\phi$, 
as was shown in \cite{AM,AFH}, 
are related through the
ratio $r= \dmsol/|\dma|= (m_2^2-m_1^2)/|m_3^2-m_1^2|$,
where $\dmsol$ and $\dma$ are neutrino mass squared differences 
driving the solar and the dominant atmospheric neutrino oscillations.
As a consequence, the Majorana CP violating phases 
depend effectively only on $\alpha$. 
The related constraints on the neutrino mass spectrum have been
studied in \cite{AM,AFH,YL}.

 In the present article we investigate in detail the generation
of baryon asymmetry of the Universe $Y_B$ within the AF and AM models.
Although our analysis is done for these two specific models,
it has generic features which are common to models based 
on $A_4$ flavour symmetry. In the class of models under 
discussion the neutrino masses arise
via the type I see-saw mechanism. 
Correspondingly, one can implement the
leptogenesis scenario of matter-antimatter asymmetry generation. As is 
well known, in this scenario the baryon asymmetry is produced through
the out of equilibrium CP violating decays  of the RH neutrinos 
$\nu^c_i$ (and their SUSY partners - RH sneutrinos $\tilde{\nu}^c_i$) 
in the early Universe \cite{FY,KRS}.
The observed value of $Y_B$ to be reproduced, given as the ratio
between the net baryon number density 
and entropy density, reads \cite{YBexp} 
\begin{equation}
Y_B\equiv \left.\frac{n_B-n_{\bar{B}}}{s} \right|_{0}= 
(8.77 \pm 0.24)\times 10^{-11}
\end{equation}
where the subscript ``0'' refers to the present epoch.

  As has been discussed in \cite{JenkinsManohar}, 
the CP asymmetries $\epsilon_i$, originating in the decays
of the RH neutrinos and sneutrinos $\nu^c_i$ and 
$\tilde{\nu}^c_i$ and relevant for the generation of the baryon 
asymmetry of the Universe, vanish at LO in the class of models 
under discussion. Thus, $\epsilon_i\neq 0$ are generated due to the NLO
corrections. These are, as already mentioned, 
different in the AF and AM models,
so that the results for $\epsilon_i$ differ in the two models. 

  We calculate the CP asymmetries $\epsilon_i$ in the
AF and AM types of models and discuss the dependence of
$Y_B$ on the parameter $\alpha$. 
This is done in versions of the two models in which 
the RH neutrinos have masses in the range 
$M_i \sim (10^{11} \div 10^{13})$ GeV. 
As discussed in \cite{AM}, the natural range
of RH neutrino masses in the AF and AM  models 
is  $M_i\sim(10^{14} \div 10^{15})$ GeV.
Such large values of the RH neutrino masses 
can lead \cite{PRST05} in SUSY theories with see-saw mechanism 
to conflict with the existing stringent experimental 
upper limits on the rates of lepton flavour violating (LFV) 
decays and reactions
\footnote{The indicated problem typically arises 
if the SUSY particles have masses in the range 
of few to several hundred GeV, accessible to the LHC 
experiments \cite{PRST05}.}, 
like $\mu \rightarrow e +\gamma$,
$\mu \rightarrow 3e$, $\mu^- + (A,Z) \rightarrow e^- + (A,Z)$, 
etc. The investigation of the LFV processes in 
the class of AF and AM models is beyond the scope of the present 
study. However, in order to avoid potential problems 
related to the issue of LFV, we work in versions of 
the AF and AM  models in which 
the scale of the RH neutrino masses is lower than 
in the original AF and AM models. This is achieved by 
minimally extending the AF and AM models through an additional $Z_2$
symmetry,  which enables us to appropriately suppress the 
neutrino Yukawa couplings. This in turn allows 
to lower the scale of RH neutrino masses down 
to $(10^{11} \div 10^{13})$ GeV.

  We perform the analysis of baryon asymmetry generation in the 
so-called ``one flavour'' approximation. The latter is valid as long as 
the masses of the RH neutrinos satisfy \cite{PPRio106,DiBGRaff06} 
$M_i \gsim 5\times 10^{11} (1 + \tan^2\beta)$ GeV, 
where $\tan\beta$ is the ratio 
of the vacuum expectation values of the two Higgs doublets present in the 
SUSY extensions of the Standard Model.
The ``one flavour'' approximation condition is satisfied
for, e.g. $\tan\beta \sim 3$ and $M_i \sim 10^{13}$ GeV. 
Actually in the models we consider relatively small values of 
$\tan\beta$ are preferable \cite{AF2}.
Further, with masses of the RH neutrinos in the range of 
$(10^{11} \div 10^{13})$ GeV one can safely neglect 
the effects of the $\Delta L=2$ wash-out processes in 
leptogenesis \cite{Giudiceetal}. This allows us to use 
simple analytic approximations in the calculation of the 
relevant efficiency factors $\eta_{ii}$.
We perform the calculation of the baryon asymmetry for the 
two types of light neutrino mass spectrum - with normal and inverted 
ordering. Both types of spectrum are allowed in 
the class of models considered.

   We find that it is possible to generate the correct size and sign 
of the baryon asymmetry $Y_B$ in the versions of both
the AF and AM models we discuss. 
The sign of $Y_B$ is uniquely determined by the sign
of $\sin \phi$, since all other factors in $Y_B$ have a definite sign.
Interestingly, in the low energy observables only $\cos \phi$ is
present, so that both $\sin \phi \lessgtr 0$ are compatible with the low
energy data.  Conversely speaking, taking into account the 
sign of the baryon asymmetry
$Y_B$ we are able to fix uniquely also the sign of $\sin\phi$,
which is otherwise undetermined through the low energy data.

   Leptogenesis is not studied for the first time in the class of
models of interest. However, our work overlaps little with 
the already existing publications on the subject.
In \cite{AdhikaryGhosal} the CP asymmetries and $Y_B$ were also calculated.
This is done, however, not within the context 
of a self-consistent model since instead of computing 
the NLO corrections, the authors introduce 
ad hoc random perturbations in the Dirac neutrino mass matrix. 
In \cite{JenkinsManohar} only the CP asymmetries are
calculated within the AF model without discussing the washout effects 
which can change the results for $Y_B$. 
In \cite{AM} results for the CP
asymmetries are also given, but the washout 
effects are not taken into account.
Finally, in \cite{leptogenesisA4other} a highly degenerate 
spectrum of masses of RH neutrinos is considered 
and the baryon asymmetry is produced via resonant leptogenesis.

   The paper is organized as follows: in Section 2 we give a short
introduction to the AF and the AM models and discuss the changes due
to adding the $Z_2$ symmetry. We also give the expressions for 
the NLO corrections relevant for the calculation of the baryon 
asymmetry $Y_B$. In Section 3 we discuss the light and heavy 
Majorana neutrino mass spectra. We study the Majorana
phases and their dependence on the parameter $\alpha$. 
Our analysis of leptogenesis in both models for 
light neutrino mass spectrum 
with normal ordering (NO) and inverted
ordering (IO)  is given in Section 4. We
summarize the results of the present work in Section 5. 
The two appendices contain details on the
flavon superpotential and the generation 
of an appropriate VEV for the additional flavon field 
$\zeta$, present in the models considered by us.

\cleqn
\mathversion{bold}
\section{Variants of the Two $A_4$ Models}
\mathversion{normal}

In this section we recapitulate the main features of the AF \cite{AF2}
and the AM model \cite{AM}. We supplement them with an additional
$Z_2$ symmetry to appropriately suppress the Dirac Yukawa couplings of
the neutrinos and to lower the mass scale of the RH
neutrinos. We explicitly check that changes in the models connected to
the $Z_2$ extension  do not affect the LO results 
for the lepton masses and mixings and also only slightly affect the NLO results. 

For an introduction to the group theory of $A_4$ we refer to
\cite{AF2,AM}, whose choice of 
generators for the $A_4$ representations we follow.

\subsection{Altarelli-Feruglio Type Model}

In this model the flavor symmetry $A_4$ is accompanied 
by the cyclic group $Z_3$ and the Froggatt-Nielsen symmetry $U(1)_{FN}$.
We add, as mentioned, a further $Z_2$ symmetry to suppress 
the Dirac couplings of the neutrinos. By assuming that the RH neutrinos
acquire a sign under $Z_2$, the renormalizable coupling becomes forbidden
\footnote{Alternatively, one could also let $h_u$ instead 
of $\nu^c$ transform under the $Z_2$ symmetry to forbid
the Dirac Yukawa coupling at the renormalizable level.}.
To allow a Yukawa coupling for neutrinos at 
all we introduce a new flavon $\zeta$  
which only transforms under $Z_2$. 
We call the VEV of $\zeta$ $\VEV{\zeta}=z$ 
in the following and assume that $z \approx \lambda_c^2 \La$
as all other flavon VEVs.
Clearly, the Majorana mass terms of the RH
neutrinos remain untouched, at LO. The symmetries and particle content of the AF variant are as given in table 1.
\begin{table}
\begin{center}
\begin{tabular}{|c||c|c|c|c|c||c||c|c|c|c|}
\hline 
\rule[0.15in]{0cm}{0cm}{\tt Field}& $l$ & $e^c$ & $\mu^c$ & $\tau^c$ & $\nu^c$ & $h_{u,d}$ & 
$\varphi_T$ & $\varphi_S$ & $\xi$, $\tilde\xi$ & $\zeta$ \\
\hline
$A_4$ & $3$ & $1$ & $1''$ & $1'$ & $3$ & $1$ & 
$3$ & $3$ & $1$ & $1$ \\
\hline
$Z_3$ & $\omega$ & $\omega^2$ & $\omega^2$ & $\omega^2$ & $\omega^2$ & $1$ &
$1$ & $\omega^2$ & $\omega^2$ & $1$ \\
\hline
$Z_2$ & $+$ & $+$ & $+$ & $+$ & $-$ & $+$ & 
$+$ & $+$ & $+$ & $-$\\
\hline
\end{tabular}
\end{center}
\begin{center}
\normalsize
\begin{minipage}[t]{15cm}
\caption[particles of the AF variant]{\scriptsize Particle content of the AF variant. 
Here we display the transformation properties of lepton
superfields, Minimal Supersymmetric SM (MSSM) Higgs and flavons under the flavor group 
$A_{4} \times Z_3 \times Z_2$. $l$ denotes the three lepton doublets, $e^c$, $\mu^c$
and $\tau^c$ are the three $SU(2)_L$ singlets and $\nu^c$ are the three RH neutrinos forming an $A_4$-triplet.
Apart from $\nu^c$ and the flavon $\zeta$ all fields are neutral under the additional $Z_2$ symmetry.
Note that $\omega$ is the third root of unity, i.e. $\omega=\mathrm{e}^{\frac{2 \pi i}{3}}$.
To accommodate the charged lepton mass hierarchy the existence a
$U(1)_{FN}$, under which only the RH charged leptons are
charged, is assumed. The $U(1)_{FN}$ is broken by
an FN field $\theta$ only charged under $U(1)_{FN}$ with charge -1.
Additionally, the model contains a $U(1)_R$ symmetry relevant for the alignment of the vacuum.
\label{tab:particles_AF} \normalsize}
\end{minipage}
\end{center}
\end{table}
At LO neutrino masses are generated by the terms\footnote{The field $\tilde\xi$ does not have a VEV at LO and thus is not relevant at this
level.}
\begin{equation}
y_\nu (\nu^c l) h_u \zeta/\Lambda + a \xi (\nu^c \nu^c) + b \, (\nu^c \nu^c \varphi_S)
\end{equation}
with $(\cdots)$ denoting the contraction to an $A_4$-invariant. Thus, the LO terms in the neutrino sector are the same as in the original
AF model, apart from the suppression of the Dirac coupling. Also the
LO result that the charged lepton mass matrix is diagonal, is not changed compared to
the original model. The mass matrices of the neutrinos are of the form ($m_D$ is given in the right-left basis)
\begin{equation}
m_D = y_\nu \left( \begin{array}{ccc}
	1 & 0 & 0\\
	0 & 0 & 1\\
	0 & 1 & 0
\end{array}
\right) \, \frac{z}{\La} v_u \;\;\; \mbox{and} \;\;\;
m_M = \left( \begin{array}{ccc}
	a u + 2 b v_S & -b v_S & -b v_S\\
	-b v_S & 2 b v_S & a u - b v_S\\	
	-b v_S & a u - b v_S & 2 b v_S
\end{array}
\right)
\end{equation}
with $v_u=\VEV{h_u}$, $\VEV{\xi}=u$ and $\VEV{\varphi_{Si}}=v_S$ according to the alignment in \cite{AF2}. The light neutrino mass matrix is
given by the type I see-saw term
\begin{equation}
m_\nu=-m_D^T m_M^{-1} m_D
\end{equation}
and has the generic size $\lambda_c^2 v_u^2/\La$. At the same time, the effective dimension-5 operator $l h_u l h_u/\La$,
\footnote{We make the ``conservative'' assumption that all non-renormalizable operators are suppressed by the same cutoff scale $\La$.}
 which can also
contribute to the light neutrino masses, is only invariant under the flavor group, if it involves two flavons of the type $\varphi_S$ and
$\xi$ ($\tilde\xi$). Thus, its contribution to the light neutrinos masses scales as $\lambda_c^4 v_u^2/\La$, which is 
always subdominant compared to the type I see-saw contribution. The size of the contribution from the effective dimension-5 operator
is actually of the same size as possible NLO corrections to the type I see-saw term.

Considering the NLO corrections note that these involve for the Dirac neutrino mass matrix either the two flavon combination $\varphi_T \zeta$
or the shift of the vacuum of $\zeta$. The first type gives rise to two different terms
\begin{equation}
y_A (\nu^c l)_{3_S} \varphi_T h_u \zeta/\Lambda^2 + y_B (\nu^c l)_{3_A} \varphi_T h_u \zeta/\Lambda^2
\end{equation}
with $(\cdots)_{3_{S \, (A)}}$ standing for the (anti-)symmetric triplet of the product $\nu^c l$.
The correction due to the shift in $\VEV{\zeta}$ 
can be simply absorbed into a redefinition of the coupling $y_\nu$. Thus, 
using the alignment of $\varphi_T$ as given in \cite{AF2},
the structure of the NLO corrections to $m_D$ is 
the same as in the original model 
\begin{equation}
\delta m_D = \left( \begin{array}{ccc}
	 2 y_A & 0 & 0\\
	 0& 0 & -y_A-y_B\\
	 0& -y_A +y_B& 0
\end{array} 
\right)\frac{v_T z}{\La^2} v_u \; .
\end{equation}
The NLO corrections to the Majorana mass matrix of the RH neutrinos are exactly the same as in the original AF model, i.e.
\begin{eqnarray}
&& a \, \delta\xi (\nu^c \nu^c) + \tilde a \, \delta\tilde\xi (\nu^c \nu^c) + b \, (\nu^c \nu^c \delta\varphi_S)\\ \nonumber
&+& x_A \, (\nu^c\nu^c) (\varphi_S \varphi_T)/\La +x_B \, (\nu^c\nu^c)' (\varphi_S \varphi_T)''/\La
+x_C \, (\nu^c\nu^c)'' (\varphi_S \varphi_T)'/\La \\ \nonumber
&+& x_D \, (\nu^c\nu^c)_{3_S} (\varphi_S \varphi_T)_{3_S}/\La +x_E \, (\nu^c\nu^c)_{3_S} (\varphi_S \varphi_T)_{3_A}/\La + x_F \, (\nu^c\nu^c)_{3_S} \varphi_T \xi/\La 
+ x_G \, (\nu^c\nu^c)_{3_S} \varphi_T \tilde\xi/\La 
\end{eqnarray}
where $\delta \varphi_S$, $\delta \xi$ and $\delta \tilde\xi$ indicate the shifted vacua of the flavons $\varphi_S$, $\xi$ and $\tilde\xi$.
Taking into account the possibility of absorbing these corrections partly into the LO result, they give rise to four independent
additional contributions to $m_M$ which can be effectively parametrized as
\begin{equation}
\delta m_M = \left( \begin{array}{ccc}
	 2 \tilde x_D & \tilde x_A & \tilde x_B-\tilde x_C\\
	 \tilde x_A &  \tilde x_B +2 \tilde x_C& -\tilde x_D\\
	 \tilde x_B-\tilde x_C & -\tilde x_D & \tilde x_A
\end{array}
\right) \; \lambda_c^4 \La \; .
\end{equation}
Compared to these, NLO corrections involving the new flavon $\zeta$ are suppressed, since invariance under the $Z_2$ symmetry
requires always an even number of $\zeta$ fields and invariance under the $Z_3$ at least one field of the type $\varphi_S$, $\xi$ or $\tilde\xi$.
The NLO corrections to the charged lepton masses are also the same as in the original model
and effects involving $\zeta$ can only arise at the level of three flavons.

As we show in appendix A, the VEV of the flavon $\zeta$ is naturally also of the order $\lambda_c^2 \Lambda$ as the VEVs
of the other flavons and the shift of its VEV is of the size $\delta \rm{VEV} \sim \lambda^2_c \rm VEV$. We also calculate its effect
on the vacuum alignment of the other flavons and show that the results achieved in the original AF model, especially the alignment at LO, 
remain unchanged.

\subsection{Altarelli-Meloni Type Model}

The AM model, proposed in \cite{AM}, possesses as flavor symmetry $A_4 \times Z_4$. To this we add a $Z_2$ symmetry under which
only the Higgs field $h_u$ and the new flavon $\zeta$ transform. Compared to the original model, we change 
the transformation properties of $h_u$ into $1''$ under $A_4$ and it transforms now trivially under $Z_4$. The flavon $\zeta$
is a $1'$ under $A_4$ and acquires a phase $i$ under $Z_4$. The transformation properties of leptonic superfields, MSSM Higgs and flavons can be found
in table 2.
\begin{table}
\begin{center}
\begin{tabular}{|c||c|c|c|c|c||c|c||c|c|c|c|c|}
\hline
{\tt Field}&  $l$ & $e^c$ & $\mu^c$ & $\tau^c$ & $\nu^c$ & $h_d$ & $h_u$& 
$\varphi_T$ &  $\xi'$ & $\varphi_S$ & $\xi$ & $\zeta$\\
\hline
$A_4$ & $3$ & $1$ & $1$ & $1$ & $3$ & $1$ & $1''$ &$3$ & $1'$ & $3$ & $1$ & $1'$\\
\hline
$Z_4$ & $i$ & $1$ & $i$ & $-1$ & $-1$ & $1$ & $1$ & $i$ & $i$  & $1$ & $1$ & $i$ \\
\hline
$Z_2$ & $+$& $+$ & $+$ & $+$ & $+$ & $+$ & $-$ & $+$ & $+$ & $+$ & $+$ & $-$\\
\hline
\end{tabular}
\end{center}
\begin{center}
\normalsize
\begin{minipage}[t]{15cm}
\caption[particles of the AM variant]{\scriptsize Particle content of the AM variant. 
Transformation properties of lepton superfields, MSSM Higgs and flavons under the flavor group 
$A_{4} \times Z_4 \times Z_2$ are shown. The nomenclature is as in table 1.
Apart from $h_u$ and the flavon $\zeta$ all fields are neutral under the additional $Z_2$ symmetry.
Apart from $A_4 \times Z_4 \times Z_2$, the model also contains a $U(1)_R$
symmetry relevant for the alignment of the vacuum, similar to the AF variant.
\label{tab:particles_AM} \normalsize}
\end{minipage}
\end{center}
\end{table}
The Dirac neutrino coupling is at LO given by
\begin{equation}
y_\nu (\nu^c l) h_u \zeta/\La \; ,
\label{ynuLOAM}
\end{equation}
which leads to the same Dirac mass 
matrix $m_D$ as in the original model, 
suppressed by the factor $\lambda_c^2$ for
$z/\La \approx \lambda_c^2$. The Majorana mass terms for the RH neutrinos
remain unaffected by the changes of the model, at LO,
\begin{equation}
M (\nu^c \nu^c) + a \xi (\nu^c \nu^c) + b \, (\nu^c \nu^c \varphi_S)
\end{equation}
so that the contribution from the type I see-saw to the 
light neutrino masses arises from
\begin{equation}
m_D = y_\nu \left( \begin{array}{ccc}
	1 & 0 & 0\\
	0 & 0 & 1\\
	0 & 1 & 0
\end{array}
\right) \, \frac{z}{\La} v_u \;\;\; \mbox{and} \;\;\;
m_M = \left( \begin{array}{ccc}
	M+ a u + 2 b v_S & -b v_S & -b v_S\\
	-b v_S & 2 b v_S & M+ a u - b v_S\\	
	-b v_S & M+ a u - b v_S & 2 b v_S
\end{array}
\right)
\label{2.10}
\end{equation}
Thereby the flavon alignment given in \cite{AM} is 
used. Equation (\ref{2.10})
leads to exact TB mixing in the neutrino sector. 
For $M \approx \lambda_c^2 \La$, as argued in \cite{AM},
we find the generic size of the light neutrino masses 
to be $\lambda_c^2 v_u^2/\La$.
The effective dimension-5 operator $l h_u l h_u/\La$ arises in our 
variant only at the two flavon level
\begin{eqnarray}\nonumber
& &  (\varphi_T \varphi_T)'' (l l) h_u^2/\La^3 +  (\varphi_T \varphi_T)' (l l)' h_u^2/\La^3 + (\varphi_T \varphi_T) (l l)'' h_u^2/\La^3 
+ ((\varphi_T \varphi_T)_{3_S} (l l)_{3_S})'' h_u^2/\La^3\\ 
& & + (\xi')^2 (l l) h_u^2/\La^3 + \xi' (\varphi_T l l)' h_u^2/\La^3  + \zeta^2 (l l) h_u^2/\La^3
\end{eqnarray}
where we omit order one couplings. Thus, its contributions to the light neutrino masses are of order 
$\lambda_c^4 v_u^2/\La$, i.e. of the same size as the expected NLO corrections to the type I see-saw contribution, and hence subdominant.

The effect of the introduction of the $Z_2$ symmetry and the new field $\zeta$ on the charged lepton sector is the following:
an insertion of three flavons, two of which are $\zeta$, gives a new LO contribution to the electron mass
\begin{equation}
\zeta^2 (e^c l \varphi_T)' h_d/\La^3 \; .
\end{equation}
Using the same alignment as in \cite{AM},
its contribution resembles the one from the 
operator with $\xi'$ instead of $\zeta$ and 
thus gives also a non-vanishing term in the (11) entry of the 
charged lepton mass matrix. The latter is of the same size as those already
encountered in the original version of the AM model.
Therefore in the variant of the AM model we are considering 
the charged lepton mass matrix is also diagonal at LO and the
correct hierarchy among the charged lepton masses is predicted.

At NLO, the Dirac couplings of the neutrinos are
\begin{equation}
y_\nu (\nu^c l) \delta \zeta h_u/\La + y_A (\nu^c l) \xi \zeta h_u/\La^2  
+ y_B (\nu^c l)_{3_S} \varphi_S \zeta h_u/\La^2 + y_C (\nu^c l)_{3_A} \varphi_S \zeta h_u/\La^2 \; .
\end{equation}
The first two contributions can be absorbed into the LO coupling $y_\nu$. Compared to the original model, the other corrections are of the same type
and generate the same structure
\begin{equation}
\delta m_D = \left( \begin{array}{ccc}
	 2 y_B & -y_B - y_C & - y_B+y_C\\
	 -y_B +y_C & 2 y_B & -y_B-y_C\\
	 -y_B-y_C & -y_B+y_C & 2 y_B
\end{array}
\right) \frac{v_S z}{\La^2} v_u \; .
\end{equation}
Note that actually the contribution associated to the coupling
$y_B$ is still compatible with TB mixing so that only $y_C$ 
can lead to deviations from the TB mixing pattern. As we will see in
section 4.2, for this reason also the CP asymmetries only depend on
the coupling $y_C$.

All effects to the Majorana mass matrix of the RH neutrinos involving $\zeta$ are negligible, since we always need at minimum 
two fields $\zeta$ and additionally have to balance the $Z_4$ charge of the operator. 
Thus, the NLO corrections are only those already present in the original model
\begin{eqnarray}
&& x_A \, (\nu^c\nu^c) \xi^2/\La +x_B \, (\nu^c\nu^c) (\varphi_S \varphi_S)/\La +x_C (\nu^c \nu^c)_{3_S} (\varphi_S \varphi_S)_{3_S}/\La
 +x_D (\nu^c \nu^c)_{3_S} \varphi_S \xi/\La\nonumber\\
&+& x_E \, (\nu^c\nu^c)' (\varphi_S \varphi_S)''/\La + x_F \, (\nu^c\nu^c)'' (\varphi_S \varphi_S)'/\La \; .
\end{eqnarray}
The first four contributions can be absorbed into the LO result (or vanish). We do not mention effects from shifts in the 
vacua of $\varphi_S$ and $\xi$, since these effects can in this model also be absorbed into the LO result. The new structures at NLO
lead to $\delta m_M$ of the form
\begin{equation}
\delta m_M = 3 \left( \begin{array}{ccc}
	 0 & x_E & x_F\\
	 x_E & x_F & 0\\
	 x_F & 0 & x_E
\end{array}
\right) \frac{v_S^2}{\La} \; .
\end{equation}
For the charged leptons, additional NLO corrections to the muon and the electron mass arise from three and four flavon insertions, respectively, 
involving the field $\zeta$.
The operator 
\begin{equation}
\zeta^2 (\mu^c l \varphi_S)' h_d/\La^3
\end{equation}
corrects the muon mass. This type of subleading contribution already exists in the original
model such that no new structures are introduced. The NLO corrections to the electron mass are induced through the operator 
$\zeta^2 (e^c l \varphi_T)' h_d/\La^3$, if the shifts in the vacua are taken into account, as well as through the four flavon operators
\begin{equation}
\zeta^2 (e^c l (\varphi_T \varphi_S)_{3_S})' h_d/\La^4 + \zeta^2 (e^c l (\varphi_T \varphi_S)_{3_A})' h_d/\La^4  
+ \zeta^2 \xi (e^c l \varphi_T)' h_d/\La^4  + \zeta^2 \xi' (e^c l \varphi_S) h_d/\La^4 \; .
\end{equation}
All structures arising from these corrections are already generated by the NLO corrections present in the original model so that 
the analysis given in \cite{AM} for the NLO corrections is still valid in the constructed variant.

In appendix B we discuss how to give a VEV of the desired size to the field $\zeta$, the shift of this VEV from NLO corrections
as well as the effects of $\zeta$ on the flavon superpotential of the original model at LO and NLO.

\cleqn
\mathversion{bold}
\section{Neutrino Masses and CP Violating Phases in the $A_4$ Models}
\mathversion{normal}

The models discussed in the previous section have in common that
the Majorana mass matrix of RH neutrinos is of the form
\begin{equation}
m_M\;=\;\left(
\begin{array}{ccc}
 X+2Z & -Z & -Z \\
 -Z & 2Z & X-Z \\
 -Z & X-Z & 2Z
\end{array}\right)\label{RH-Mass}
\end{equation}
and the neutrino Dirac mass matrix reads
\begin{equation}
m_D\;=\;y_\nu\left(
\begin{array}{ccc}
 1 & 0 & 0 \\
 0 & 0 & 1 \\
 0 & 1 & 0
\end{array}\right)\frac{z}{\La}v_u
\end{equation}
The symmetry of this class of models implies that, at leading
order, the neutrino part of the Lagrangian depends only 
on few parameters: $X,Z$ and $y_\nu$. These
parameters are, in general, complex numbers. One
can set $y_\nu$ real by performing a global phase transformation of the
lepton doublet fields. 
As we will see,  CP violating phases, 
which enter in the CP asymmetries of the RH neutrino
decays, are functions of the relative phase between $X$ and $Z$.
The see-saw mechanism for the neutrino mass generation implies that 
the full parameter space of the neutrino sector 
can be constrained significantly by the low energy data.

The RH neutrino mass matrix (\ref{RH-Mass}) is
diagonalized by an orthogonal matrix $U_{TB}$, given in (\ref{eq:UTB}): 
\begin{equation}
 \diag(M_1 e^{i\f_1},M_2 e^{i\f_2},M_3 e^{i\f_3})=U_{TB}^T m_M U_{TB}\,, 
\end{equation}
where
\begin{eqnarray}
M_1 &=& |X+3Z| \;\equiv\;|X|\,|1+\alpha e^{i\phi}|,\,\,\,\,\f_1={\rm arg}(X+3Z)  \label{M1}\\
M_2 &=& |X|,\,\,\,\,\,\f_2={\rm arg}(X) \label{M2}\\
M_3 &=& |X-3Z| \;\equiv\;|X|\,|1-\alpha e^{i\phi}|,\,\,\,\,\,\f_3={\rm arg}(3Z-X). \label{M3}
\end{eqnarray}
Here $\alpha\equiv|3Z/X|$ and $\phi\equiv{\rm arg}(Z)-{\rm
arg}(X)$.

A light neutrino Majorana mass term is generated after electroweak
symmetry breaking via the type I see-saw mechanism:
\begin{equation}\label{mnuLO}
m_\nu  \;=\;- m_D^T\, m_M^{-1}\, m_D=U^*\diag
\left(m_1,m_2,m_3\right)U^\dagger
\end{equation}
where 
\begin{equation}
U=i\,U_{TB}\,\diag\left(e^{i\f_1/2},e^{i\f_2/2},
e^{i\f_3/2}\right) \label{U_matrix}
\end{equation}
and $m_{1,2,3}$ are the light neutrino masses,
\begin{equation}
 m_i\equiv\frac{(y_\nu)^2
v_u^2}{M_i}\left(\frac{z}{\La}\right)^2\,,\,\,\,\,\,i=1,2,3\label{LO_masses}
\end{equation}
The $i$ in eq. (\ref{U_matrix}) correspond to an unphysical common phase and we will
ignore it in what follows.
 We observe also that
one of the phases $\f_k$, say $\f_1$, can be considered as a common phase of the neutrino mixing matrix, 
and thus has no physical relevance. In the following we always set $\f_1=0$.

The parameters $|X|$, $\alpha$ and $\phi$ defined in
(\ref{M1})-(\ref{M3}), which determine the RH
neutrino mass matrix (\ref{RH-Mass}), can be constrained by the neutrino oscillation
data. More specifically, we have for the ratio \cite{AM}:
\begin{equation}
 r\equiv \frac{\dmsol}{|\dma|}\;=\;\frac{(1+\alpha^2-2\,\alpha\,\cos \phi)(\alpha+2\cos \phi)}{4\,|\cos
 \phi|}\,,\label{ratio_mass}
\end{equation}
where $\dmsol=\dmB\equiv m^2_2-m^2_1>0$ and $|\dma|=|\dmC|\cong|\dmD|$ are the $\nu-$mass squared differences responsible respectively for solar and atmospheric neutrino oscillations. Since the value of $r$ is fixed by the data, 
this relation implies a strong correlation 
between the values of the parameters  $\alpha$ 
and $\cos\phi$. Let us note that the sign of $\sin\phi$ cannot be constrained by the low energy data.
As we will see later,
the sign of $\sin\phi$ is fixed by the sign of the baryon asymmetry of the Universe, computed
in the leptogenesis scenario.

At $3\sigma$, the following experimental constraints must be
satisfied \cite{nudata1}:
\begin{eqnarray}
\dmsol &>& 0 \nn \\
|\dma| &=& (2.41 \pm 0.34)\times 10^{-3}\, \eV^2 \label{condnorm} \\
r&=& 0.032 \pm 0.006\nn~.
\end{eqnarray}

\begin{figure}[t!!]
\begin{center}
\includegraphics[width=8.5cm,height=6.5cm]{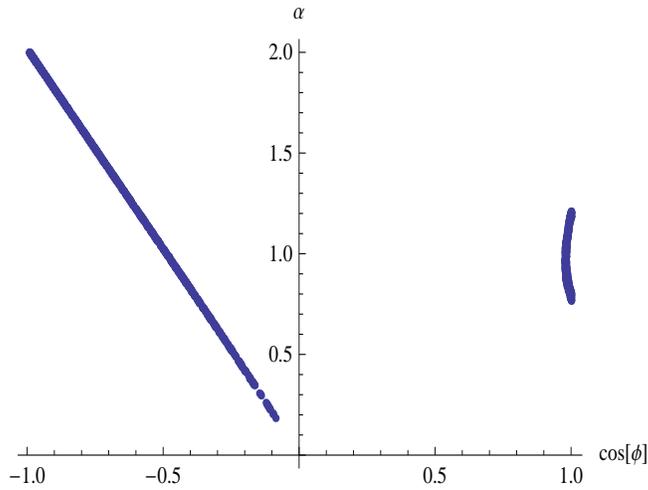}
\caption{The correlation between the real parameter $\alpha$ and the phase 
$\phi$, which appear in the RH neutrino Majorana mass matrix. 
The figure is obtained by using the $3\sigma$ range of 
the parameter $r$ given in eq. (\ref{ratio_mass}).
See text for details.
\label{Fig.1}}
\end{center}
\end{figure}

In Fig. \ref{Fig.1} we show the correlation between $\alpha$ and $\cos\phi$,
following from (\ref{ratio_mass}) taking into account (\ref{condnorm}).
Depending on the sign of $\cos\phi$, the parameter space is divided into two physically distinctive parts: 
$\cos\phi>0$ corresponds to light neutrino mass spectrum with
normal ordering (NO), whereas for $\cos\phi<0$ one obtains neutrino mass spectrum with inverted 
ordering (IO) .

 The main difference between the models we are discussing 
and the original models
reported in \cite{AF2}  and
\cite{AM} is in the mass scale of the RH
neutrino fields. In the models considered here the predicted RH
neutrino masses are always rescaled by the additional 
factor $(\lambda_c^2)^2 \sim 10^{-3}$.
Depending on the
value of the neutrino Yukawa coupling $y_\nu$, in our model
the lightest RH Majorana neutrino mass can be in the range from
$(10^{11} \div 10^{12})$ GeV, and up to $10^{13}$ GeV 
for a neutrino Yukawa coupling $y_\nu\sim\mathcal{O}(1)$.

 For neutrino mass spectrum with NO, the RH neutrino masses
show approximately the following partial hierarchy \cite{JenkinsManohar}:
$M_1\approx 2 M_2 \approx 10 M_3$. The lightest neutrino mass
$m_1$, compatible with the experimental constraints given in
(\ref{condnorm}), takes values in the interval $3.8\times 10^{-3}
{\rm eV}\lesssim m_1 \lesssim 6.9\times 10^{-3} {\rm eV}$. This implies that 
the light neutrino mass spectrum is with partial hierarchy 
\footnote{This was noticed also in \cite{leptogenesisS4}.}. 
For the sum of the neutrino masses we have:
\begin{eqnarray}
6.25\times 10^{-2}~{\rm eV} \lesssim m_1 + m_2 + m_3 
\lesssim 6.76\times 10^{-2}~{\rm eV}\,.
\end{eqnarray}
%
 
 In the case of IO spectrum, 
the overall range of variability of
the lightest neutrino mass, $m_3$, is the following: 
$0.02\, {\rm eV} \lesssim m_3 \leq 0.50\, {\rm eV}$, 
where only the lower bound follows from the 
low energy constraints. The upper bound was chosen by us
to be compatible with the ``conservative'' 
cosmological upper limit on the sum of the neutrino masses 
\cite{Fuk06,WMAPsummi}. Thus, the light neutrino mass spectrum 
can be with partial hierarchy or quasidegenerate.  
If the spectrum is with partial hierarchy (i.e. 
$0.02\, {\rm eV} \lesssim m_3 < 0.10~{\rm eV}$), 
for the RH Majorana neutrino masses, 
to a good approximation, we have:
$M_1\cong M_2 \cong M_3/3$.  
Quasidegenerate light neutrino mass spectrum 
implies that, up to corrections $\sim\mathcal{O}(r)$,
one has $M_1 \cong M_2 \cong M_3$.
The sum of the light neutrino masses in the case of IO spectrum 
is predicted to satisfy: 
\begin{eqnarray}
m_1 + m_2 + m_3 
\gsim 0.125~{\rm eV}\,.
\end{eqnarray}
%

   We give below the expressions for the lightest neutrino 
mass in the NO and IO spectrum
as functions of $\alpha$ and $r$. Recall that for fixed
value of $r$, all the parameter space and the associated
phenomenology is characterized by the parameter
$\alpha$. In the numerical examples reported in the following, we
always use the best fit value of the ratio $r$: $r=0.032$. By expanding with
respect to $r$, we get for the square of the lightest neutrino mass: 
\begin{equation}
 m^2_1\;=\;\dma r\left( \frac{1}{1+2\a^2}+
\frac{2(1+\a^2)r}{(1+2\a^2)^3}\right)\,,\,\,\,\,{\rm NO~ spectrum}\;;
\label{m1norm}
\end{equation}
\begin{equation}
 m^2_3\;=|\dma|\left( \frac{1}{2\a^2}+
\frac{(1+\a^2)r}{\a^2(1+2\a^2)} \right)\;,\,\,\,\,\,{\rm IO~spectrum}\;.
\label{m3inv}
\end{equation}
For $\alpha = 1$ the expression for $m^2_1$ reduces 
to the one obtained in \cite{AM}.

   In the class of models we are
considering, the three light neutrino masses obey 
the general sum rule (valid for both types of spectrum) \cite{AM,AFH}:
\begin{equation}
\frac{e^{i\f_3}}{m_3}\;=\;\frac{1}{m_1}-\frac{2e^{i\f_2}}{m_2}\label{sumrule}
\end{equation}

This equation implies a strong correlation between the neutrino
masses and the Majorana phases arising from the RH neutrino mass
matrix. The Majorana phases are responsible for CP
violation in leptogenesis and therefore we will discuss them in
detail in the following subsection.

\subsection{The Majorana CP Violating Phases}

In the following, we use the standard parametrization
of the PMNS matrix (see, e.g. \cite{Bilenky:2001rz,STPNu04}):
 \begin{equation}
 \label{eq:Upara}
  U_{\rm PMNS}\;=\;\left( \begin{array}{ccc}
 c_{12} c_{13} & s_{12} c_{13} & s_{13}e^{-i \delta}  \\[0.2cm]
  -s_{12} c_{23} - c_{12} s_{23} s_{13} e^{i \delta}
  & c_{12} c_{23} - s_{12} s_{23} s_{13} e^{i \delta}
 & s_{23} c_{13}  \\[0.2cm]
  s_{12} s_{23} - c_{12} c_{23} s_{13} e^{i \delta} &
  - c_{12} s_{23} - s_{12} c_{23} s_{13} e^{i \delta}
  & c_{23} c_{13} \\
                \end{array}
    \right)
 ~{\rm diag}(1, e^{i \frac{\alpha_{21}}{2}}, e^{i \frac{\alpha_{31}}{2}})
 \end{equation}\newline
where $c_{ij}\equiv\cos\theta_{ij}$,
$s_{ij}\equiv\sin\theta_{ij}$, $\theta_{ij}\in[0,\pi/2]$,
$\delta\in[0,2\pi]$ is the Dirac CP violating phase and
$\alpha_{21}$ and $\alpha_{31}$ are the two Majorana CP violating
phases, $\alpha_{21,31}\in[0,2\pi]$. From the see-saw mass formula
(\ref{mnuLO}) one can read directly the form of the neutrino
mixing matrix that arises at leading order in perturbation theory.
Taking into account the standard parametrization (\ref{eq:Upara})
and (\ref{eq:UTB}), the PMNS
matrix is indeed:
\begin{equation}
 U_{\PMNS}=\diag(1,1,-1)U_{TB}\,\diag(1,e^{i\f_2/2},e^{i\f_3/2}) 
\label{UPMNS}
\end{equation}
From eqs. (\ref{eq:Upara}) and (\ref{UPMNS}) we identify the 
``low energy'' Majorana phases as 
\begin{eqnarray}
 \alpha_{21}=\f_2\\
 \alpha_{31}=\f_3
\end{eqnarray}
 We remark that, at this order of perturbation
theory, the CHOOZ mixing angle, $\theta_{13}$, is always zero
as a consequence of the TB form of the neutrino mixing
matrix, imposed by the broken $A_4$ discrete symmetry.

  In the models under discussion the Majorana phases 
$\alpha_{21}$ and $\alpha_{31}$ can
also be constrained by using the neutrino oscillation data 
\footnote{Let us recall that the probabilities of 
oscillations involving the flavour neutrinos do not depend on
the Majorana phases of the PMNS matrix \cite{BHP80,Lang87}. Thus, 
the Majorana phases cannot be {\it directly} constrained by 
the neutrino oscillation data.}. 
After some algebraic
manipulation, we arrive at the following relations between the CP
violating phases $\a_{21}$ and $\a_{31}$ and the parameters 
$\alpha$ and $\phi$ of the model:
\begin{eqnarray}
 \tan\alpha_{21} &=& -\,\frac{\a\sin\phi}{1+\a\cos\phi}\label{alpha21}\\
 \tan\alpha_{31} &=& 2\,\frac{\a\sin\phi}{\a^2-1}\,,\label{alpha31}
\end{eqnarray}
where $\alpha$ and $\cos\phi$ satisfy
eq. (\ref{ratio_mass}).

In the case of NO spectrum, we have $\phi=0\pm\e,
2\pi\pm\e$, with $\e<0.2$ and $0.8\lesssim\alpha\lesssim 1.2$ (see
Fig. 1). If $\e\cong 0$, the two CP violating phases become 
unphysical. No CP violation is possible in this case.
As regards the IO light neutrino mass spectrum, 
it is easy to show that \cite{AM} $2\cos\phi\approx -\a$. 
The correction, $\delta_\alpha(\alpha)$, 
which appears in the right-hand side of this equation, is given by
\begin{equation}
\delta_\alpha(\alpha)\;=\;\frac{2\alpha
r}{1+2\alpha^2}\left(1-\frac{2(1+\alpha^2)r}{(1+2\alpha^2)^2}\right)
\end{equation}

For light neutrino mass spectrum with inverted ordering, the parameter $\alpha$
varies in the interval $0.07\lesssim\alpha\lesssim 2$, where the
lower limit of $\alpha$ comes from the 
indicative upper bound on the absolute neutrino mass scale 
used by us, $m_{1,2,3}\lesssim0.5$ eV.

\begin{figure}[t!!]
\begin{center}
\begin{tabular}{cc}
\includegraphics[width=7.2cm,height=5.5cm]{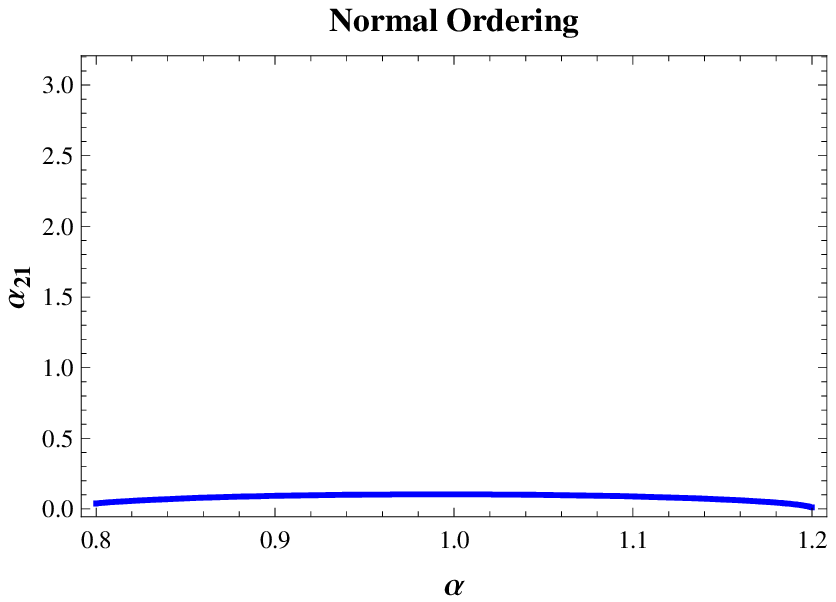} &
\includegraphics[width=7.2cm,height=5.5cm]{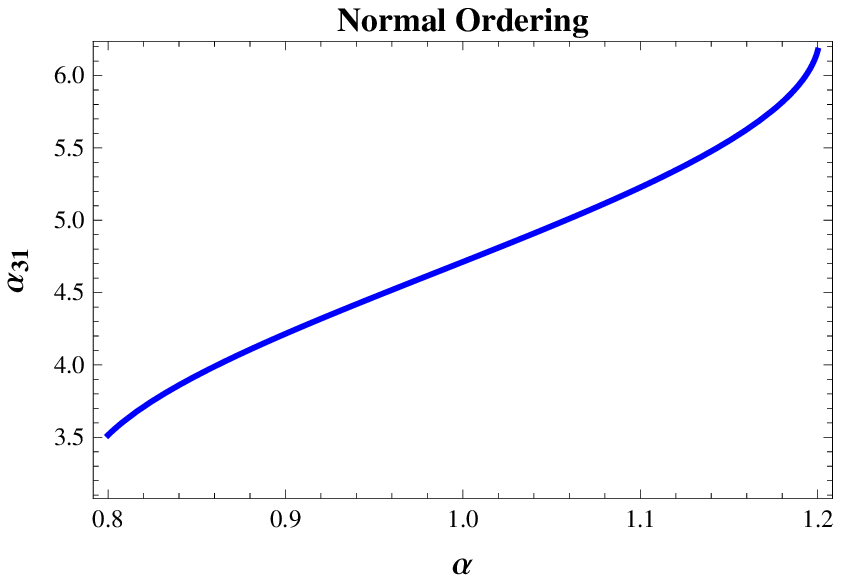}
\end{tabular}
\caption{The Majorana phases $\alpha_{21}$ and $\alpha_{31}$
in the case of a light neutrino mass spectrum with normal ordering. 
The parameter $r$ is set to its best fit value, $r=0.032$.
The solutions of equations (\ref{alpha21}) and (\ref{alpha31}) 
shown in the figure correspond to $\sin\phi<0$. See text for details.}
\end{center}
\label{PhNO}
\end{figure}

\begin{figure}[t!!]
\begin{center}
\begin{tabular}{cc}
\includegraphics[width=7.2cm,height=5.5cm]{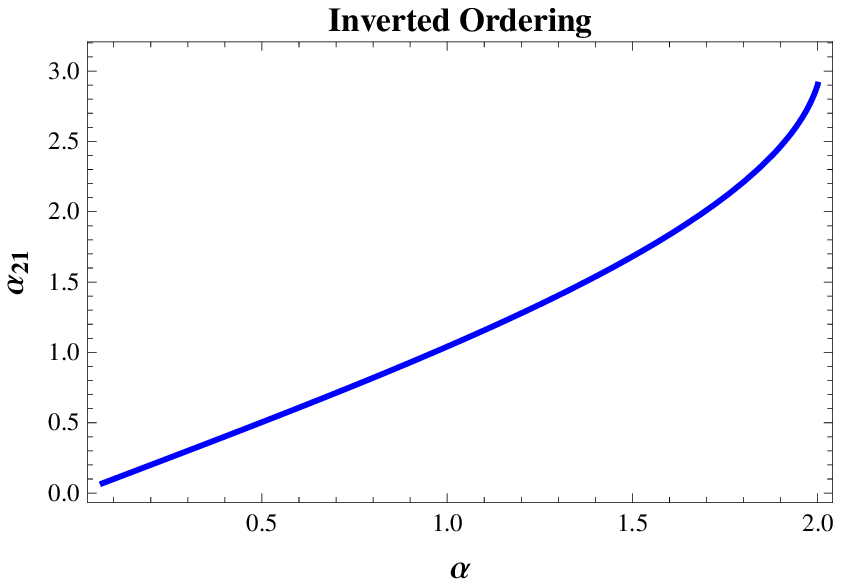} &
\includegraphics[width=7.2cm,height=5.5cm]{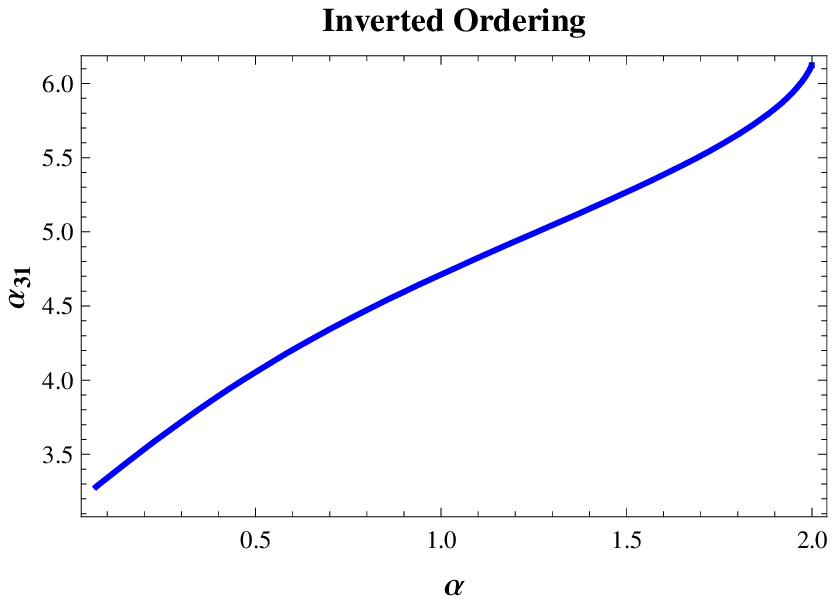}
\end{tabular}
\caption{The same as in Fig. 2, but for a light neutrino 
mass spectrum with inverted ordering.
}
\end{center}
\end{figure}

In Figs 2 and 3 we show the behavior of the Majorana phases
$\alpha_{21}$ and $\alpha_{31}$ as functions of 
$\alpha$, for the NO and IO mass spectrum, respectively.
We choose $\sin\phi<0$ in (\ref{alpha21}) and (\ref{alpha31}).
As we will see, this is dictated by reproducing correctly
the sign of the baryon asymmetry.
 On the other hand, the relative sign of 
$\sin\a_{21}$ and $\sin\a_{31}$ is fixed by the requirement that the
sum rule in eq. (\ref{sumrule}) is satisfied. In the case of NO 
spectrum, the phase $\alpha_{21}$ is close to zero.
The maximum value of $\alpha_{21}$ is obtained for $\alpha\cong1$. 
At $\alpha=1$ we have approximately
 $\alpha_{21}=\sqrt{r/3}\cong 0.1$.
The other Majorana phase $\alpha_{31}$ can assume large CP
violating values. The largest $|\sin\a_{31}|$ is reached for 
$\alpha=1$: at this value of $\alpha$ we have $\sin\a_{31}=-1$. 

If the light neutrino mass spectrum is with IO,
both phases can have large CP violating values. 
We get $\sin\a_{21}=1$ and $\sin\a_{31}=-1$
for $\alpha\approx\sqrt{2}$ and $\alpha=1$, respectively.

The Majorana phase $\alpha_{21}$ can be probed,
in principle, in the next generation of experiments searching for
neutrinoless double beta decay \cite{Nextbb0nu}. 
Below we give the expression of the effective Majorana mass
$m_{ee}$ predicted in the class of models we are considering in the cases
of neutrino mass spectrum with normal and inverted ordering 
\cite{Bilenky:2001rz}: 
\begin{eqnarray}
m_{ee} &\cong& \left|m_1\cos^2\theta_{12}+
\sqrt{m_1^2+\dmsol}\sin^2\theta_{12} e^{i\a_{21}}\right|\,,\,\,\,\,\,{\rm NO}\,;\\
m_{ee} &\cong& \sqrt{m_3^2+|\dma|}\left|\cos^2\theta_{12} 
+ e^{i\alpha_{21}}\sin^2\theta_{12}\right|\,,\,\,\,\,\,{\rm IO}\;.\\\nonumber
\end{eqnarray}
%
We recall that in the class of models under discussion, 
$\sin^2\theta_{12} = 1/3$, $\cos^2\theta_{12} = 2/3$ and  
a non-zero value of $\theta_{13}$ arises only due to the NLO
corrections. As a consequence, 
the predicted value of $\theta_{13}$ is relatively 
small, $\theta_{13}\sim\mathcal{O}(\lambda_c^2\sim 0.04)$. 
Thus, the terms $\sim\sin^2\theta_{13}$ in $m_{ee}$
give a negligible contribution.
Further, since the Majorana phase $\a_{21}\cong 0$ (see Fig. 2),
the two terms in the expression for $m_{ee}$ in the 
case of NO spectrum add up. As a consequence, we have:
\begin{eqnarray}
m_{ee} &\cong& \left|\frac{2}{3}\,m_1 + \frac{1}{3}\,
\sqrt{m_1^2+\dmsol}\right|\,,\,\,\,\,\,{\rm NO}\,,
 \end{eqnarray}
%
where  
$3.8\times 10^{-3}~{\rm eV} \lesssim m_1 \lesssim 
6.9\times 10^{-3}~{\rm eV}$. 

\begin{figure}[t!!]
\begin{center}
\includegraphics[width=8.5cm,height=6.5cm]{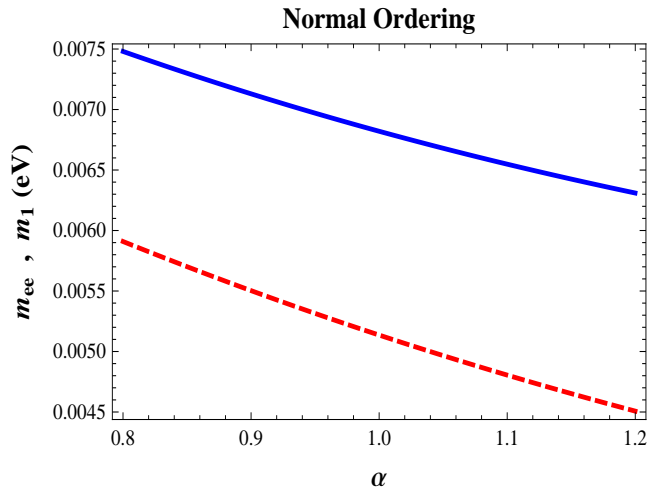} 
\caption{The effective Majorana mass $m_{ee}$ (blue continuous line) 
and lightest neutrino mass $m_1$ (red dashed line) 
in the case of a light neutrino mass spectrum with normal
ordering. In both cases $\dma$ and $r$ are fixed 
to their best fit values.
\label{fig3}}
\end{center}
\end{figure}

We show in Fig. \ref{fig3} the effective Majorana 
mass $m_{ee}$ and the lightest neutrino mass 
$m_1$ versus $\alpha$, for the NO spectrum.  
In this case, $m_{ee}$ takes values in the interval: 
$6.5\times 10^{-3}\,{\rm eV} \lesssim m_{ee} \lesssim 
7.5\times 10^{-3}\,{\rm eV}$. 
Similar conclusion has been reached in \cite{leptogenesisS4}.
In what concerns the IO spectrum, the full range 
of variability of the effective Majorana mass, compatible with 
neutrino oscillation data, is predicted to be 
\begin{eqnarray}
\frac{1}{3}\,\sqrt{m_3^2+|\dma|} 
\lesssim m_{ee} \lesssim \sqrt{m_3^2+|\dma|}\,,\,\,{\rm with}~~
m_3 \gsim 0.02~{\rm eV}\,,\,\,\,{\rm IO}\;.
\end{eqnarray}
%
For $m_3 \gsim 0.02~{\rm eV}$, this implies 
$m_{ee}\gsim 0.018\,\,{\rm eV}$~~(see also \cite{AM}).

\cleqn
\mathversion{bold}
\section{Leptogenesis}
\mathversion{normal}

In this section we compute the baryon asymmetry within the
AF and AM type models defined in Section 2.
As we have already noticed earlier, leptogenesis cannot be
realized if we take into account only the leading order
contribution to the neutrino superpotential. In order to generate a
sufficiently large CP asymmetry, higher
order corrections to the Dirac mass matrix of neutrinos
must be taken into account. The RH neutrino mass spectrum in 
this class of models is not strongly
hierarchical. Consequently, the standard thermal leptogenesis
scenario in which the  
relevant lepton CP violating asymmetry is generated 
in the decays of the lightest RH (s)neutrino only is 
not applicable
and one has to take into account 
the contribution 
from the out of equilibrium decays 
of the heavier RH (s)neutrinos. 
The lepton asymmetry thus produced
in the decays of all heavy RH (s)neutrinos  
$\nu^c_i$ ($\tilde{\nu}^c_i$ ), $i=1,2,3$, 
is converted into a baryon number by sphaleron 
interactions. The neutrino and sneutrino
CP asymmetry $\epsilon_i$, which are equal for lepton 
and slepton final states,
is the following \cite{Covi:1996wh}:
\begin{equation}
 \epsilon_i\;=\;\frac{1}{8\pi v_u^2}
\sum_{j\neq i}\frac{{\rm Im}[(\hat{m}_D \hat{m}_D^\dagger)^2_{ji}]}
{(\hat{m}_D\hat{m}_D^\dagger)_{ii}}\;
f\left(\frac{m_i}{m_j} \right)\;,
\label{eps}
\end{equation}
where 
\begin{equation}
\hat{m}_D=U^\dagger m_D\label{hat_mD} 
\end{equation}
is the neutrino Dirac mass matrix in the 
mass eigenstate basis 
of RH neutrinos, 
and $m_i$, $i=1,2,3$
are the light neutrino masses.
The  matrix $U$ and the masses $m_i$ 
coincide with those given in 
eqs. (\ref{U_matrix}) and (\ref{LO_masses}), 
respectively. 
The loop function $f(m_i/m_j)$ is defined as
\begin{equation}
 f(x)\;\equiv\;
 -x\left(\frac{2}{x^2-1}+\log\left(1+\frac{1}{x^2}\right)\right)\label{f}
\end{equation}
%
This function depends strongly on the hierarchy of
light neutrino masses. It can lead to a strong enhancement 
of the CP asymmetries if the light neutrino 
masses $m_i$ and $m_j$  are nearly degenerate. 
As we have seen earlier,
the neutrinos can be quasidegenerate in mass 
in the case of IO  spectrum. 
In this case we have to a good approximation
$f(m_i/m_j)\cong -f(m_j/m_i)$. 

 We recall that in the case of IO spectrum, the lightest 
two heavy Majorana (s)neutrinos, $\nu^c_{1,2}$ ($\tilde{\nu}^c_{1,2}$ ), 
have very close masses.
However, the conditions for resonant leptogenesis 
\cite{ResLG}
are not satisfied in the models under consideration.
Indeed, in all the region of the relevant parameter space
we have $0.2\lesssim\alpha\lesssim2$. Correspondingly,
the relative mass difference of the two 
heavy Majorana (s)neutrinos in question is
\begin{equation}
\left|\frac{M_2-M_1}{M_1}\right|=1-\frac{m_1}{m_2}\cong(2 \div 14)\times
10^{-3}\gg {\rm
max}\left|\frac{(\hat{m}_D\hat{m}_D^\dagger)_{12}}{16\pi^2
v_u^2}\right|\approx \frac{\lambda_c^6}{\pi^2}\approx {10^{-5}}\,
\end{equation}
%
Under the above condition, the CP asymmetries
for each (s)neutrino decay can be computed in perturbation theory as
the interference between the tree level and one loop diagrams 
(see, e.g. \cite{Buchmuller:1997yu,Blanchet:2006dq}). 

 The general expression for the baryon asymmetry \cite{HT90}, 
in which each RH (s)neutrino gives a non-negligible contribution, 
can be cast in the following form:
\begin{equation}
 Y_B\equiv\frac{n_B -\bar{n}_B}{s}\;
=\; -1.48\times 10^{-3} \sum_{i,j=1}^3 \epsilon_i \eta_{ij}
\end{equation}
where $\eta_{ij}$ is an efficiency factor that accounts for the
effects of washout due to the $\D L=1$ 
interactions of $\n^c_j$ and $\tilde{\n}^c_j$
of the asymmetry $Y_{l_i}$, generated in the decays 
$\n^c_i\rightarrow l_i\,h_u,\,\tilde{l}_i\,\tilde{h}_u$ and
$\tilde{\nu}^c_i\rightarrow \tilde{l}_i\,h_u,\,l_i\, \tilde{h}_u$. 
They take into account also the decoherence effects 
on $l_i$ caused by the $\n^c_j$ and $\tilde{\n}^c_j$ ($j\neq i$) 
interactions. We refer in the following
discussion only to the lepton number densities. The same
considerations apply for the interactions involving slepton
states.

 The computation of the efficiency factors in the models under discussion
is considerably simplified \cite{Engelhard:2006yg,Engelhard:2007kf} 
(see also \cite{Nielsen:2002pc}). This is due to the fact that 
to leading order, the heavy Majorana neutrinos 
$\n^c_1$, $\n^c_2$ and $\n^c_3$, as can be shown,
couple to orthogonal leptonic states. As a consequence, the 
Boltzmann evolutions of the three lepton CP violating  
asymmetries, associated with the indicated three 
orthogonal leptonic states, are practically independent.
Taking into account the above considerations, one can
compute the total baryon asymmetry
as an incoherent sum of the contributions arising 
from decays of each of the three heavy RH neutrinos:
\begin{equation}
 Y_B\approx\sum_{i=1}^3 Y_{Bi}\;,
\end{equation}
where 
\begin{equation}
 Y_{Bi}\;\equiv\;-1.48\times 10^{-3}\epsilon_i\,\eta_{ii} \label{YBi}\;.
\end{equation}

  In the class of models considered 
the RH neutrino mass scale is set
below $10^{14}$ GeV, preventing possible washout effects from $\D
L=2$ scattering processes. 
In this case, the efficiency factors $\eta_{ii}$
can be expressed only in terms of the washout mass parameters
$\mtil_i$ \cite{Giudiceetal}:
\begin{equation}
 \eta_{ii}\;=\;\left(\frac{3.3\times 10^{-3}\,{\rm eV}}{\mtil_i}+\left(\frac{\mtil_i}{0.55\times 10^{-3}\,{\rm
 eV}}\right)^{1.16}\right)^{-1}\label{eta}\;,
\end{equation}
where 
\begin{equation}
\label{mtil}
 \mtil_i\;\equiv\;\frac{(\hat{m}_D \hat{m}_D^\dagger)_{ii}}{M_i}\;.
\end{equation}
Here $\hat{m}_D$ is the neutrino Dirac mass matrix 
in the basis in which the Majorana mass matrix of RH neutrinos 
is diagonal with real eigenvalues (see eq. (\ref{hat_mD}).

\subsection{Leptogenesis in the Variant of AF Model}

In this Section we compute the baryon asymmetry for the
AF type model, in the one-flavor leptogenesis
regime.

 In the basis in which the RH Majorana neutrino mass term given 
in (2.2) is diagonal,
the relevant matrix that enters into the expression of the
leptogenesis CP asymmetries (\ref{eps}) is given by
 \begin{eqnarray}
  &&\hat{m}_D\hat{m}_D^\dagger \;=\; \bold{1}\,\left(\frac{z}{\Lambda}\right)^2 y^2_\nu v^2_u\nn \\\\ 
   &&\;+\;
\left(
 \begin{array}{ccc}
  2 \,{\rm Re}(y_A)& 2\sqrt{2}e^{i\frac{\a_{21}}{2}}\,{\rm Re}(y_{A})& \frac{2}{\sqrt{3}} e^{i\frac{\a_{31}}{2}}\,{\rm Re}(y_B) \\
   2\sqrt{2}e^{-i\frac{\a_{21}}{2}}\,{\rm Re}(y_{A}) & 0 &  -2 \sqrt{\frac{2}{3}}e^{i\frac{\a_{31}-\a_{21}}{2}}\,{\rm Re}(y_B) \\
   \frac{2}{\sqrt{3}} e^{-i\frac{\a_{31}}{2}}\,{\rm Re}(y_B)&  -2 \sqrt{\frac{2}{3}} e^{i\frac{\a_{21}-\a_{31}}{2}}\,{\rm Re}(y_B)& -2 \,{\rm Re}(y_A)
 \end{array}\right)\left(\frac{v_T}{\La}\right)\left(\frac{z}{\Lambda}\right)^2 y_\nu v^2_u \nn
 \end{eqnarray}
\noindent where $y_A$ and $y_B$  are the higher order 
(complex) Yukawa couplings defined in (2.4) and (2.5).
We can take all flavon VEVs real without loss of generality.\newline

The CP asymmetries $\epsilon_k$ ($k=1,2,3$) can be
written in the following way:
\begin{equation}
\epsilon_1 =  -\frac{1}{6\pi}\left(\frac{z}{\La}\right)^2\left(\frac{v_T}{\La}\right)^2\left(6 f(m_1/m_2)\sin\alpha_{21}\,{\rm Re}(y_{A})^2+
 f(m_1/m_3)\sin\alpha_{31}\,{\rm Re}(y_{B})^2 \right)\label{e1AF}
\end{equation}
\begin{equation}
 \epsilon_2 = \frac{1}{3\pi}\left(\frac{z}{\La}\right)^2\left(\frac{v_T}{\La}\right)^2\left(3 f(m_2/m_1)\sin\alpha_{21}\,{\rm Re}(y_{A})^2+
 f(m_2/m_3)\sin(\alpha_{21}-\alpha_{31})\,{\rm Re}(y_{B})^2 \right)
\end{equation}
\begin{equation}
 \epsilon_3 = \frac{1}{6\pi}\left(\frac{z}{\La}\right)^2\left(\frac{v_T}{\La}\right)^2\left( 2 f(m_3/m_2)
 \sin(\alpha_{31}-\alpha_{21}) + f(m_3/m_1)\sin\alpha_{31}\right)\,{\rm Re}(y_{B})^2\label{e3AF}
\end{equation}
where $m_{k}$ are the LO neutrino masses and 
$z/\La\approx v_T/\La\approx\lambda_c^2$. Thus, in the model
under consideration we have
\begin{equation}
 |\epsilon_k|\;\propto\;\lambda_c^8\,\approx\,6\times 10^{-6}\,,\,\,\,\,\,k=1,2,3\;.
\end{equation}

This is the order of magnitude we expect 
for the CP asymmetry if we require successful leptogenesis. 
Depending on the loop factor $f(m_i/m_j)$ (eq. \ref{f}) and the 
values of the Majorana phases, the CP asymmetry can be 
enhanced or suppressed.

The washout mass parameters, associated to each of the three 
lepton asymmetries, are given by:
\begin{eqnarray}
\mtil_1 &=& m_1 (1+\mathcal{O}(\lambda_c^2)) \label{mtil1}\\
\mtil_2 &=& m_2 (1+\mathcal{O}(\lambda_c^2)) \\
\mtil_3 &=& m_3 (1+\mathcal{O}(\lambda_c^2)) \label{mtil3}
\end{eqnarray}
We see that the washout mass parameters, 
to a good approximation, coincide with the neutrino masses.

\paragraph{Results for NO Spectrum\\\\}

We study the baryon asymmetry in the region of the parameter space
corresponding to a neutrino mass spectrum with normal ordering:
$0.8\lesssim\alpha\lesssim 1.2$. The lightest RH Majorana
neutrino in this scenario is $\n^c_3$. The Majorana
phases, that provide the requisite CP violation 
for a successful leptogenesis, are solutions of equations
(\ref{alpha21}) and (\ref{alpha31}) corresponding to $\sin\phi<0$.
The dependence on $\alpha$ of each of the two CP violating phases
is shown in Fig. 2. We recall that only the solutions corresponding 
to $\sin\phi<0$ give the correct sign of the total
baryon asymmetry.

 We show in Fig. 5, left panel, the dependence 
of the baryon asymmetry $Y_B$ on the parameter $\alpha$. 
The individual contributions to $Y_B$ from the decays of each of the three 
RH Majorana neutrinos are also shown. 
The term $Y_{B3}$, originating from the lightest
RH neutrino decays, is suppressed by largest washout
effects, with respect to $Y_{B1}$ and $Y_{B2}$ (see
(\ref{mtil3})). 

The contribution to the total baryon asymmetry given by $Y_{B1}$
shows an interplay between two independent terms
proportional to $y_A$ and $y_B$, respectively. These two terms 
have always the same signs and are of the same order of magnitude. 
The suppression due to the Majorana phase
$\alpha_{21}\lesssim 0.1$ of the term proportional to $y_A$ is
compensated by the enhancement due to the loop factor: we find that
$6f(m_1/m_2)/f(m_1/m_3)\cong - (8 \div 20)$. 
The same considerations apply to $Y_{B2}$. 
Now $\sin\alpha_{21}$
and $\sin(\alpha_{21}-\alpha_{31})$ have the same sign
and the ratio of the corresponding loop factors 
is approximately
$3 f(m_2/m_1)/f(m_2/m_3)\cong (20 \div 30$).

In conclusion, in the case of NO light neutrino mass
spectrum, each of the two Majorana phases $\alpha_{21}$ and
$\alpha_{31}$, having values within the ranges 
allowed by neutrino oscillation data (see Fig. 2), 
can provide the CP violation which is required 
in order to have
successful leptogenesis. 
Even in the case in which the term
proportional to $\sin{\alpha_{31}}$ in the CP asymmetries is
strongly suppressed (which corresponds to the case of 
strong fine-tuning of $y_B\ll 1$), successful baryogenesis 
can be naturally realised for values of the Majorana 
phase $\alpha_{21}\approx (0.04 \div 0.10)$ and a moderately large 
neutrino Yukawa coupling 
$y_A\sim (2.5 \div 3.0)$.

\begin{figure}[t!!]
\begin{center}
\begin{tabular}{cc}
\includegraphics[width=7.2cm,height=5.5cm]{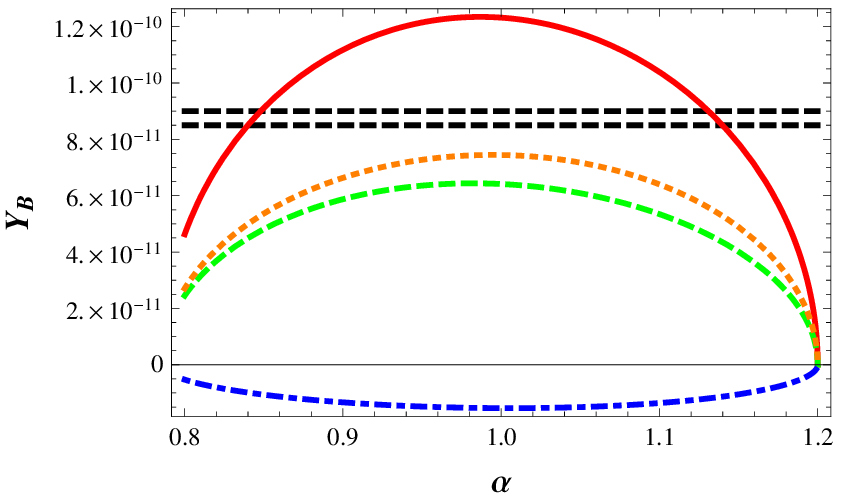} &
\includegraphics[width=7.2cm,height=5.5cm]{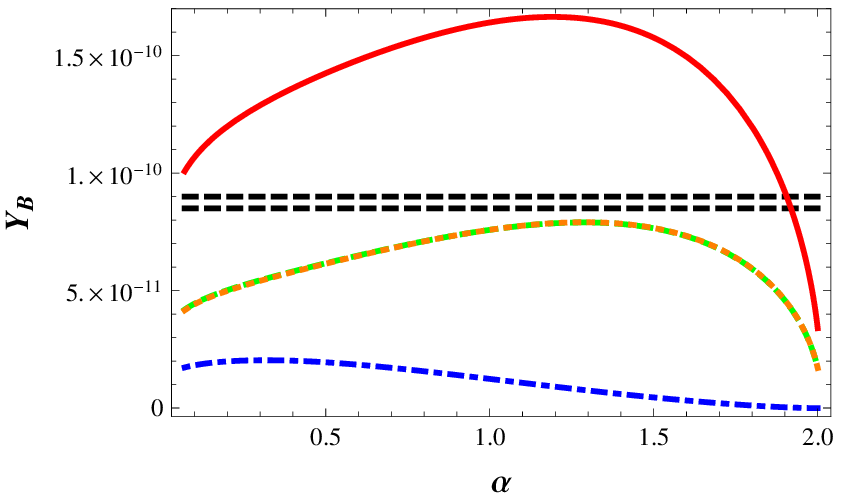}
\end{tabular}
\caption{AF type model: baryon asymmetry versus
$\alpha$ in the cases of neutrino mass spectrum with normal 
(left panel) and inverted (right panel)
ordering. In each plot we show: $i)$ the
total baryon asymmetry $Y_B$ (red continuous curve), $ii)$
$Y_{B1}$ (green dashed
curve), $iii)$ $Y_{B2}$ (orange dotted curve) and $iv)$ $Y_{B3}$ 
(blue dot-dashed curve). On the right panel, the 
lines corresponding to $Y_{B1}$ and $Y_{B2}$ overlap. 
In both cases $\sin\phi<0$ and $\dma$ and $r$ are fixed 
at their best fit values. The results shown 
in the left (right) panel correspond to
$y_A=2.5$ and $y_B=3$ ($y_A=0.4$ and $y_B=2$). 
The horizontal dashed lines represent the allowed range of 
the observed value of $Y_B$, $Y_B\in[8.5,9]\times 10^{-11}$.  
\label{Fig5}}
\end{center}
\end{figure}

\paragraph{Results for IO Spectrum\\\\}

We now study in detail the region of the parameter space for which
the neutrino mass spectrum is with inverted ordering and is 
hierarchical. This scenario
is realized for $0.2 <\alpha\lesssim 2$. In the
following, we report the behavior of the baryon asymmetry in all
the interval of variability of $\alpha$, compatible with an
IO neutrino spectrum and 
for which the computation of the CP asymmetry can be done 
in perturbation theory. 
Thus, the results we show for 
$0.07 <\alpha\lesssim 0.2$ should be valid provided the 
renormalisation group (RG) effects \cite{RGrunningnus} 
are sufficiently small in the indicated region.

In Fig. \ref{Fig5}, right panel, we plot the different contributions to the
baryon asymmetry, as we have done previously for the normal
hierarchical mass spectrum.

The Majorana CP violating phases which enter into the expressions for the 
CP asymmetries are reported in Fig. 3. 
The solutions of equations
(\ref{alpha21}) and (\ref{alpha31}) corresponding to 
$\sin\phi<0$ must be used also in this case  
in order to obtain the correct sign of the baryon
asymmetry. Now $\nu^c_3$ is the heaviest RH Majorana 
neutrino and the washout effects for the CP asymmetry 
generated in the decays of this state are less strong
since they are controlled to LO by the lightest 
neutrino mass $m_3$: $\mtil_3 = m_3$.
We note, however, that also in this scenario 
the contribution of the term $Y_{B3}$ in $Y_B$ 
is always much smaller than the 
contribution of the other two terms $Y_{B1}$ and $Y_{B2}$.
This is a consequence of the strong 
enhancement in the self energy part of the loop function that
enters into the expressions for $Y_{B1}$ and $Y_{B2}$. 
Indeed, if the spectrum is inverted hierarchical,
we have $f(m_1/m_2)\cong-f(m_2/m_1)\approx 50 f(m_3,m_{1,2})$. 
For this reason the CP violating phase $\alpha_{31}$
gives, in general, a subdominant contribution in the CP asymmetries
$\epsilon_1$ and $\epsilon_2$ when the Yukawa couplings $y_A$ and $y_B$ 
are of the same order of magnitude. This conclusion is valid
even in the region of the parameter space where 
$\alpha_{31}\approx 3\pi/2$.

The analysis of all the parameter space defined by $\alpha$,
compatible with low energy neutrino oscillation data, in the
AF type model, shows that in both the normal and
inverted patterns of light neutrino masses, the Majorana phases
can provide enough CP violation in order to have successful
leptogenesis, even in the case in which only one of the phases
$\a_{21}$ and $\a_{31}$, effectively, contributes in the
generation of the CP asymmetry.

\subsection{Leptogenesis in the Variant of AM Model}

\begin{figure}[t!!]
\begin{center}
\begin{tabular}{cc}
\includegraphics[width=7.2cm,height=5.5cm]{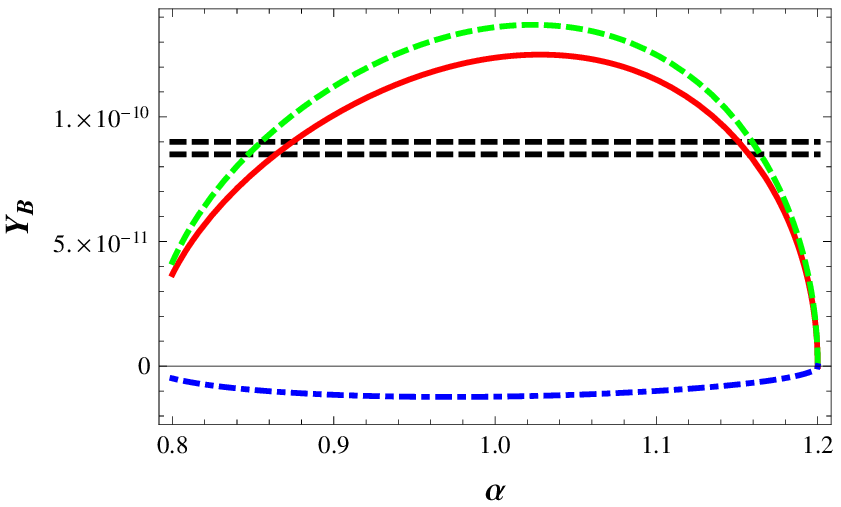} &
\includegraphics[width=7.2cm,height=5.5cm]{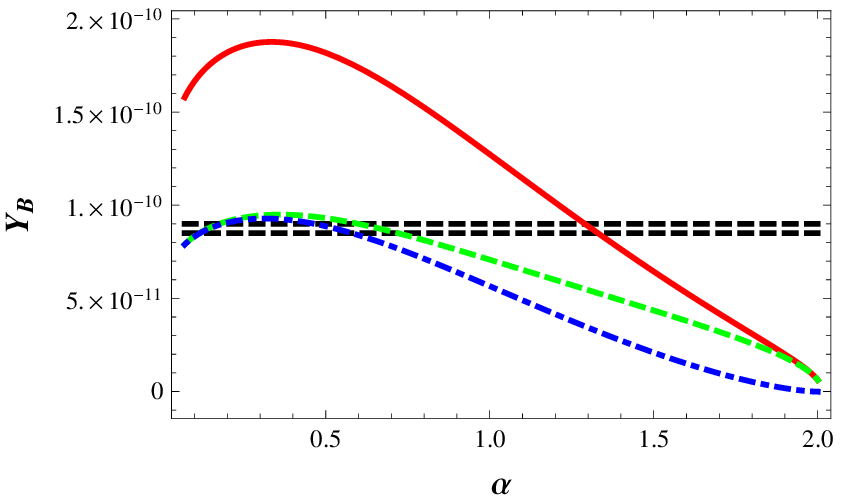}
\end{tabular}
\caption{The same as in Fig. 5 but for the AM type model.
The Yukawa coupling $y_C$ is set to $y_C=2$ in both the figures.\label{Fig6} }
\end{center}
\end{figure}

In this section we study the generation of the baryon asymmetry of
the Universe in the variant of the AM model considered by us.
We work in the one-flavour leptogenesis approximation.
The quantity relevant for the calculation 
of the CP asymmetries in this case is:
 \begin{eqnarray}
  &&\hat{m}_D\hat{m}_D^\dagger \;=\; \bold{1}\,\left(\frac{z}{\La}\right)^2 y^2_\nu v^2_u\nn \\\\ 
	&&\;+\;
\left(
 \begin{array}{ccc}
  6 \,{\rm Re}(y_B)& 0 & 2\sqrt{3}e^{i\frac{\a_{31}}{2}}\,{\rm Re}(y_C) \\
  0 & 0 & 0 \\
  2\sqrt{3}e^{-i\frac{\a_{31}}{2}}\,{\rm Re}(y_C) & 0 & -6 \,{\rm Re}(y_B)
 \end{array}\right)\left(\frac{v_S}{\La}\right)\left(\frac{z}{\La}\right)^2 y_\nu v^2_u\nn
 \end{eqnarray}
\noindent where the $y_B$ and $y_C$ are defined in eqs. (2.13) and
(2.14). Again we can choose all flavon VEVs to be 
real without loss of generality.\newline

  The CP asymmetries in this model are given by
\begin{eqnarray}
\epsilon_1 &=& -\frac{3}{2\pi}\left(\frac{z}{\La}\right)^2\left(\frac{v_S}{\La}\right)^2
 f(m_1/m_3)\sin(\a_{31})\,{\rm Re}(y_C)^2\\
 \epsilon_2 &=& 0\\
 \epsilon_3 &=& \frac{3}{2\pi}\left(\frac{z}{\La}\right)^2\left(\frac{v_S}{\La}\right)^2 f(m_3/m_1)\sin(\a_{31})\,{\rm Re}(y_C)^2
\end{eqnarray}
where $m_{1,3}$ are again the LO neutrino masses (see eq. (\ref{LO_masses})). 
The leptogenesis CP violating phase now
coincides with the Majorana phase $\alpha_{31}$.
Moreover, the CP asymmetries $\epsilon_{1,3}\neq 0$ 
are controlled by only one parameter,
$y_C$, of the matrix of neutrino Yukawa couplings  (2.14), 
the reason being that  only this parameter breaks the TB form of 
the latter.

  As we see, the heavy RH Majorana neutrino $\nu^c_2$ ``decouples'':
the CP violating lepton asymmetry is produced in the out of
equilibrium decays of the heavy Majorana neutrinos 
$\nu^c_1$ and $\nu^c_3$ alone. This constitutes a major
difference  with the variant of the AF 
model, analyzed in the preceding subsection.
After the lepton asymmetries are 
converted into a baryon asymmetry
by sphaleron processes,
the final matter-antimatter asymmetry of the
Universe can be estimated as:
\begin{equation}
 Y_B\equiv Y_{B1}+Y_{B3}
\end{equation}
where $Y_{Bi}$, for $i=1,3$, are given in eq. (\ref{YBi}). 
The LO washout mass parameters $\mtil_{1,3}$ are the
same as in the variant of the AF model:
\begin{eqnarray}
\mtil_1 &=& m_1(1+ \mathcal{O}(\lambda_c^2))\\
\mtil_3 &=& m_3(1+ \mathcal{O}(\lambda_c^2))
\end{eqnarray}

 In Fig. \ref{Fig6} we show the dependence of the baryon
asymmetry on the parameter $\alpha$ in the cases of 
neutrino mass spectrum with 
normal and inverted ordering. Both types of spectrum 
are allowed in the model considered.
The ranges of possible values of the 
Majorana phase $\alpha_{31}$ which provides the
correct sign of the baryon asymmetry are 
shown for the NO and IO spectra in Figs. 2 and 3
(they are the same as for the AF type model).

   We observe that, as in the variant of the AF model, the
suppression of the term $Y_{B3}$ with respect to $Y_{B1}$ 
in the case of NO spectrum is due to 
the relatively larger washout effects 
in the generation of the asymmetry $\epsilon_{3}$. 
The maximum of the total baryon asymmetry $Y_B$ 
is reached for $\alpha\approx 1$ where the
CP violating Majorana phase $\alpha_{31} \cong 3\pi/2$.
(see Fig. 2, right panel).

  In what concerns the IO spectrum, 
the two terms $Y_{B1}$ and $Y_{B3}$ enter with
the same sign in the total baryon asymmetry and are of the same
order of magnitude. 
The enhancement of the asymmetry for $\alpha< 0.7$ 
is explained by the increase of the loop function
$f(m_1/m_3)\cong-f(m_3/m_1)$ in the region of quasi-degenerate
light neutrino mass spectrum.

In this class of models, successful leptogenesis can be naturally
realized for both types of spectrum - NO and IO,
for an effective Yukawa coupling $y_C\gsim 1.5$.

\section{Summary}

  In the present work we studied the related issues 
of Majorana CP violating phases and 
leptogenesis in variants of two 
prominent (and rather generic)
supersymmetric $A_4$ models \cite{AF2,AM} 
which naturally lead at leading order 
(LO) to tri-bimaximal (TB) mixing in the lepton sector. 
The pattern of neutrino mixing  
suggested by the existing neutrino oscillation 
data is remarkably similar to the TB one.
Both models are supersymmetric and 
employ the type I see-saw mechanism 
of neutrino mass generation.
They predict at LO a diagonal mass matrix 
for charged leptons and lead to exact TB mixing in the neutrino
sector. The mass matrix of the RH neutrinos contains only two
complex parameters $X$, $Z$.
All low energy observables are expressed through
only three independent quantities: the real parameter 
$\alpha=|3 Z/X|$, the relative
phase $\phi$ between $X$ and $Z$, and the absolute scale of the light 
neutrino masses. The latter is a combination of the neutrino
Yukawa coupling  and the parameter $|X|$ which determines 
the scale of RH neutrino masses. 

 The main difference between the original models and those 
considered by us
is in the scale of RH neutrino masses. 
In the original models this scale is around 
$(10^{14} \div 10^{15})$ GeV.
In order to avoid possible potential problems with LFV  processes
we consider versions of both models 
in which the scale of RH neutrino masses is lower, namely, 
is in the range of $(10^{11} \div 10^{13})$ GeV. 
This is achieved by imposing an additional $Z_2$ symmetry 
capable of suppressing sufficiently the neutrino
Yukawa couplings. As a consequence, 
the mass scale of the RH neutrinos is lowered as well. 
We discussed in detail the flavon superpotential in 
the modified models of interest.
The results obtained at leading order and 
next to leading order (NLO) 
in the original models are still valid 
in the extensions  we consider. 

  The two Majorana phases of the PMNS matrix, 
$\alpha_{21}$ and $\alpha_{31}$,
effectively play the role of leptogenesis 
CP violating parameters in the generation of the 
baryon asymmetry. In the models considered both the 
phases $\alpha_{21}$ and $\alpha_{31}$ 
and the ratio $r \equiv \dmsol/|\dma|$
are functions of the two parameters 
$\alpha$ and $\phi$. We analyzed in detail the dependence of the two
``low energy'' Majorana phases $\alpha_{21}$ and $\alpha_{31}$
on $\alpha$ and $\phi$.  In contrast to the low energy observables, 
like neutrino masses and the 
effective Majorana mass in neutrinoless double beta decay, $m_{ee}$,
we show that these phases depend both on $\sin \phi$ and $\cos \phi$, 
and not only on $\cos \phi$.  
We show also that the sign of the baryon asymmetry $Y_B$ 
uniquely determines the sign of $\sin \phi$, which has 
to be negative: $\sin \phi < 0$.

   In the case of neutrino mass 
spectrum with normal ordering (NO),
$\alpha_{21}$ is shown to be small, $\alpha_{21} \lesssim 0.1$. 
In the types of models considered $\sin^2\theta_{13}$ is also 
predicted to be small,  $\sin^2\theta_{13} \sim 10^{-3}$.
As a consequence, the contributions of the terms 
$\propto \sin^2\theta_{13}$ in $m_{ee}$ are strongly suppressed.
The lightest neutrino mass is predicted to lie in the interval
$(3.8 \div  6.9)\times 10^{-3}$ eV, thus 
the neutrino mass spectrum is with partial hierarchy.
The effective Majorana mass  $m_{ee}$ has a relatively large value, 
$m_{ee}\sim 7 \times 10^{-3}$ eV. We note that if 
$\alpha_{21}$ had a value close to $\pi$, one would have 
$m_{ee} \ll 10^{-3}$ eV.
Depending on $\alpha$, the phase $\alpha_{31}$  
can take large CP violating values.  
For light neutrino mass spectrum with inverted ordering (IO),
the Majorana CP phases $\alpha_{21}$ and
$\alpha_{31}$ vary (for $\sin \phi <0$)
between $0$ and $\pi$ and $\pi$ and $2 \pi$,
respectively. 

Throughout this study we have neglected 
renormalization group effects
on neutrino masses and mixings which 
can be large for a quasi-degenerate (QD) light 
neutrino mass spectrum. 
A QD spectrum can arise in the
models considered if  $\dma < 0$ (i.e. 
the spectrum is with inverted ordering) and 
$\alpha \lesssim 0.2$. However, this corresponds only to a
small portion of the parameter space of the models.
 
 As has been already discussed in the literature, 
in the models of interest leptogenesis is not
possible at LO:
the corresponding CP asymmetries $\epsilon_i$ vanish. 
Thus, the inclusion of NLO effects is crucial for the
generation of the baryon asymmetry $Y_B$. More precisely, 
the NLO effects correcting the neutrino Dirac mass matrix 
$m_D$ give rise to non-vanishing $\epsilon_i$ and 
therefore to non-vanishing $Y_B$. Due to this fact the 
CP asymmetries are  naturally of the order of $10^{-6}$ 
(independent of the precise value of the loop function). 
Further, although the AF and  AM type
models lead to the same results at LO, they differ at NLO so that the
CP asymmetries generated in the two models are different. 

  We find that it is possible to generate the correct size and sign 
of the baryon asymmetry $Y_B$ in the versions of both
the AF and AM models we discuss. 
The study of leptogenesis was performed in 
the framework of the one flavor
approximation and by using analytic formulae for the 
relevant efficiency factors $\eta_{ii}$. 
Since the mass spectrum of the RH neutrinos is generically not
strongly hierarchical, the decays of all three RH (s)neutrinos 
contribute to the generation of the baryon asymmetry.
We find that the correct magnitude as well as the correct sign of the
baryon asymmetry $Y_B$ can be easily obtained in the AF and AM type
models for most values of the parameter $\alpha$ and natural values
of the NLO couplings. As already mentioned, the sign of $Y_B$ 
uniquely fixes the  sign of  $\sin\phi$. 
The latter cannot be determined by low energy observables 
since they exhibit only a $\cos\phi$-dependence. 

  To conclude, the results of our detailed study show 
that SUSY models with $A_4$ flavour symmetry and type 
I see-saw mechanism of neutrino mass
generation, which gives rise to tri-bimaximal mixing 
and Majorana CP violation in the lepton sector, can account 
also successfully for the observed baryon asymmetry of the Universe.

\section*{Acknowledgements}
We thank Werner Rodejohann and Yasutaka Takanishi for
discussions. C.H. would like to thank the Max-Planck-Institut f\"{u}r
Kernphysik for kind hospitality during the last stages of the work.
This work was supported in part by the Italian INFN 
under the program ``Fisica Astroparticellare''
and by the European Network ``UniverseNet'' (MRTN-CT-2006-035863).

\appendix

\newpage
\cleqn
\section{Flavon Superpotential in the AF Type Model}

In the construction of the flavon superpotential we closely follow \cite{AF2} and introduce an additional
$U(1)_R$ symmetry under which driving fields have charge +2, superfields containing SM fermions +1 and flavons, $h_{u,d}$
and FN field(s) are uncharged. To give a VEV of order $\lambda_c ^2 \Lambda$
to $\zeta$ we introduce a new driving field $\zeta_0$ which is a singlet under all symmetries of the model, apart
from carrying a $U(1)_R$ charge +2. The terms contributing to the flavon superpotential containing $\zeta_0$ at LO
read \footnote{Terms such as $\zeta_0 h_u h_d$ are not relevant, since we assume that the flavor symmetry is broken much above the electroweak scale.}
\begin{equation}
w^{\zeta}_d = M_\zeta^2 \zeta_0 + g_a \zeta_0 \zeta^2 + g_b \zeta_0
(\varphi_T \varphi_T) \; .
\end{equation}
Analogously to the original model, we demand a vanishing $F-$ term of $\zeta_0$
\begin{equation}
M_\zeta^2 + g_a \zeta^2 + g_b (\varphi_{T1}^2 + 2 \varphi_{T2}
\varphi_{T3}) =0 \; .
\end{equation}
At the same time, the field $\zeta$ does not couple to the other driving fields, $\varphi^T_0 \sim (3,1)$, $\varphi^S_0 \sim (3,\omega^2)$
and $\xi_0 \sim (1,\omega^2)$ under $(A_4,Z_3)$, in the model at LO. Thus, their $F-$terms
read as in \cite{AF2}. We find as solution
\begin{equation}
z^2 = - \frac{1}{g_a} \left( M_\zeta^2 + g_b v_T ^2 \right)
\end{equation}
and the same results for the VEVs of $\varphi_T$, $\varphi_S$, $\xi$ and $\tilde\xi$ as in \cite{AF2}.
For the mass parameter $M_\zeta$ being of order $\lambda_c^2 \Lambda$ the VEV $z$ is also of order $\lambda_c^2 \Lambda$.

Concerning the NLO contributions stemming from $\zeta$ to the alignment of the flavons $\varphi_T$, $\varphi_S$,
$\xi$ and $\tilde\xi$ we find just one term
\begin{equation}
\frac{t_z}{\Lambda} \, \zeta^2 \left( \varphi^T_0 \varphi_T \right) 
\end{equation}
which gives an additional contribution 
\begin{equation}
\frac{3 t_z}{2 g g_a} \left( g_b + \frac{M_\zeta^2}{v_T^2} \right) \frac{v_T^2}{\Lambda} 
\end{equation}
to the shift $\delta v_{T1}$ of $\varphi_T$. Its size is $\lambda_c^4 \Lambda$, as expected.
Furthermore, the shifts $\delta v_{T2,3}$ remain unchanged and thus still equal. The shifts in the vacuum of $\varphi_S$ and $\tilde \xi$
are also unchanged and the VEV of $\xi$ is still a free parameter. 

The NLO terms affecting $w^{\zeta}_d$ read
\begin{equation}
\Delta  w^{\zeta}_d = \frac{1}{\Lambda} \sum \limits _{i=1} ^{8} z_i I^Z_i
\end{equation}
with
\begin{equation}
\begin{array}{llll}
I_1^Z = \zeta_0 (\varphi_T \varphi_T \varphi_T) \; , & \;\;\;\;\;
I_2^Z = \zeta_0 (\varphi_S \varphi_S \varphi_S)  \; ,  \;\;\;\;\;
I_3^Z = \zeta_0 \xi (\varphi_S \varphi_S)  \; ,  \;\;\;\;\;
I_4^Z = \zeta_0 \tilde\xi (\varphi_S \varphi_S)  \; ,  \;\;\;\;\; \\ 
I_5^Z = \zeta_0 \xi^3 \; , & \;\;\;\;\; 
I_6^Z = \zeta_0 \xi^2 \tilde\xi \; ,  \;\;\;\;\; \;\;\;\;\; \;\;\;\;\;
I_7^Z = \zeta_0 \xi \tilde\xi^2 \; ,  \;\;\;\;\; \;\;\;\;\; \;\;\;\;
I_8^Z = \zeta_0 \tilde\xi^3 \; . 
\end{array}
\end{equation}
The result for the shift in the VEV of $\zeta$, $z + \delta z$, in the usual linear approximation, is
\begin{eqnarray}\nonumber
\delta z &=& \frac{g_b \tilde g_4}{2 g \tilde g_3 g_a} \left( t_{11} + \frac{\tilde g_4 ^2}{3 \tilde g_3 ^2} (t_6+t_7+t_8) \right) \frac{u^3}{z \Lambda}
-  \frac{3 g_b t_z}{2 g g_a^2} \left( g_b - \frac{g_a t_3}{t_z} + \frac{M_\zeta^2}{v_T^2} \right) \frac{v_T^3}{z \Lambda}\\
& & - \frac{1}{2 g_a} \left( z_1 \left(\frac{v_T^3}{u^3}\right) + \frac{\tilde g_4^2}{3 \tilde g_3 ^2} z_3 + z_5 \right) \frac{u^3}{z \Lambda}
\end{eqnarray}
with $g_4 = -\tilde g_4^2$ and $g_3 = 3 \tilde g_3^2$ as introduced in \cite{AF2}.
This shift $\delta z$ in $\VEV{\zeta}$ is of order $\lambda_c ^4 \Lambda$. Additionally, we find that - unless some non-trivial
relation among the couplings in the flavon superpotential is fulfilled - the VEVs of all driving fields vanish at the minimum.

\cleqn
\section{Flavon Superpotential in the AM Type Model}

In order to induce a VEV for the flavon $\zeta$ we add a driving field $\zeta_0$ which transforms as $1'$ under $A_4$, with $-1$ under $Z_4$ and is 
invariant under the $Z_2$ symmetry. Since it is responsible for the vacuum alignment, its charge under the $U(1)_R$ symmetry is +2. 
The LO potential for $\zeta_0$
is of the form
\begin{equation}
w^{\zeta}_d = g_a \zeta_0 \zeta^2 + g_b \zeta_0 (\varphi_T
\varphi_T)'' + g_c \zeta_0 (\xi')^2 \; .
\end{equation}
From the $F$-term of $\zeta_0$ we can derive
\begin{equation}
g_a \zeta^2 + g_b (\varphi_{T2}^2 + 2 \varphi_{T1} \varphi_{T3}) + g_c
(\xi')^2 =0 \; .
\end{equation}
Thus, $z$ takes the value
\begin{equation}
z^2 = -\frac{1}{g_a} \left( g_b v_T^2 + g_c (u')^2 \right) = -\frac{1}{g_a} \left( \frac{g_b h_1^2}{4 h_2^2} + g_c  \right) (u')^2
\end{equation}
so that $z \propto u'$ holds in case of no accidental cancellations. $u'$ is a free parameter in \cite{AM} which is taken to be of order
$\lambda_c^2 \La$.

As one can check, the field $\zeta$ does not have renormalizable interactions with the driving fields, $\varphi^T_0 \sim (3,-1)$,
$\varphi^S_0 \sim (3,1)$ and $\xi_0 \sim (1,1)$ under $(A_4,Z_4)$, of the original model. Thus, the results for the vacuum alignment found
in \cite{AM} still hold.

At NLO the field $\zeta$ contributes to the flavon superpotential of the original model through
\begin{equation}
\frac{1}{\La}\zeta^2 (\varphi^T_0 \varphi_S)' \; ,
\end{equation}
while it does not introduce any contribution at this level involving $\varphi^S_0$ or $\xi_0$. 

The NLO effects on the vacuum alignment of the field $\zeta$ stem from (order one coefficients are omitted)
\begin{eqnarray}
\frac{1}{\La} \zeta_0 \zeta^2 \xi + \frac{1}{\La} \zeta_0 (\varphi_T \varphi_T \varphi_S)'' + \frac{1}{\La} \zeta_0 \xi (\varphi_T \varphi_T)''
+ \frac{1}{\La} \zeta_0 \xi' (\varphi_T \varphi_S)' + \frac{1}{\La} \zeta_0 \xi' \xi' \xi \; .
\end{eqnarray}
Computing the effect of all NLO terms on the vacuum alignment one finds that still all shifts $\delta v_{Si}$ are equal, i.e. the shifts
do not change the structure of the vacuum, that the generic size of
all shifts - for mass parameters and VEVs of order $\lambda_c^2 \La$ -
is $\lambda_c^4 \La$ and the free parameter $u'$ is still undetermined.

Eventually, we checked that all driving fields can have a vanishing VEV at the minimum.


\end{document}